\documentclass[journal]{IEEEtran}

\UseRawInputEncoding

\usepackage{hyperref}
\hypersetup{hypertex=true,
            colorlinks=true,
            linkcolor=blue,
            anchorcolor=blue,
            citecolor=blue}
            
\usepackage{mathtools}

\usepackage{cite}

\ifCLASSINFOpdf
  \usepackage[pdftex]{graphicx}
\else
\fi

\usepackage{amsmath, amsthm}
\usepackage{amsmath, amsfonts}
\newtheorem{theorem}{Theorem}
\newtheorem{proposition}{Proposition}
\newtheorem{remark}{Remark}

\usepackage{verbatim}
\usepackage{algorithmic}
\usepackage[ruled, linesnumbered]{algorithm2e}

\usepackage{caption}
\usepackage{cleveref}

\usepackage{pdfpages}

\usepackage{array}

\usepackage{graphicx}
\usepackage{stfloats}
\usepackage{enumitem}

\usepackage{ulem}

\ifCLASSOPTIONcompsoc
  \usepackage[caption=false,font=normalsize,labelfont=sf,textfont=sf]{subfig}
\else
  \usepackage[caption=false,font=footnotesize]{subfig}
\fi

\usepackage{amssymb}

\usepackage{multicol}
\usepackage{multirow}

\newcommand{\tabincell}[2]{\begin{tabular}{@{}#1@{}}#2\end{tabular}}

\begin{document}

\title{Efficient Beam Selection for ISAC in Cell-Free Massive MIMO via Digital Twin-Assisted \\Deep Reinforcement Learning}

\author{
   Jiexin~Zhang,
   ~\IEEEmembership{Graduate Student Member,~IEEE,}
   Shu~Xu,
   ~\IEEEmembership{Member,~IEEE,}\\
   Chunguo~Li,
   ~\IEEEmembership{Senior Member,~IEEE,}
   Yongming~Huang,
   ~\IEEEmembership{Fellow,~IEEE,}
   and Luxi~Yang,
   ~\IEEEmembership{Senior Member,~IEEE}

  \thanks{
  	 
    J. Zhang, C. Li, Y. Huang, and L. Yang are with the National Mobile Communications Research Laboratory, Frontiers Science Center for Mobile Information Communication and Security, Southeast University, Nanjing 210096, China, and also with the Pervasive Communications Center, Purple Mountain Laboratories, Nanjing 211111, China  (e-mail: jiexinz@seu.edu.cn; chunguoli@seu.edu.cn; huangym@seu.edu.cn; lxyang@seu.edu.cn).
    
    S. Xu is with the School of Electronic and Optical Engineering, Nanjing University of Science and Technology, Nanjing 210094, China, and also with the National Mobile Communications Research Laboratory, Southeast University, Nanjing 210096, China (e-mail: shuxu@njust.edu.cn).}

}

\maketitle
\begin{abstract}
Beamforming enhances signal strength and quality by focusing energy in specific directions. This capability is particularly crucial in cell-free integrated sensing and communication (ISAC) systems, where multiple distributed access points (APs) collaborate to provide both communication and sensing services. In this work, we first derive the distribution of joint target detection probabilities across multiple receiving APs under false alarm rate constraints, and then formulate the beam selection procedure as a Markov decision process (MDP). We establish a deep reinforcement learning (DRL) framework, in which reward shaping and sinusoidal embedding are introduced to facilitate agent learning. To eliminate the high costs and associated risks of real-time agent-environment interactions, we further propose a novel digital twin (DT)-assisted offline DRL approach. Different from traditional online DRL, a conditional generative adversarial network (cGAN)-based DT module, operating as a replica of the real world, is meticulously designed to generate virtual state-action transition pairs and enrich data diversity, enabling offline adjustment of the agent’s policy. Additionally, we address the out-of-distribution issue by incorporating an extra penalty term into the loss function design. The convergency of agent-DT interaction and the upper bound of the Q-error function are theoretically derived. Numerical results demonstrate the remarkable performance of our proposed approach, which significantly reduces online interaction overhead while maintaining effective beam selection across diverse conditions including strict false alarm control, low signal-to-noise ratios, and high target velocities.  
\end{abstract}


\IEEEpeerreviewmaketitle

\section{Introduction}
Integrated sensing and communication (ISAC) represents a key advancement for future sixth-generation (6G) networks, facilitating the simultaneous operation of communication and radar sensing within the same spectral resources \cite{9737357,10403776}. This dual functionality not only supports high-data-rate applications but also enhances spectral efficiency \cite{9540344,elfiatoure2024multiple}. ISAC systems within cellular networks have attracted significant attention, particularly in beamforming design \cite{10681491,9791349}, parameter estimation \cite{10859257}, and resource allocation \cite{9805471,10891254}, typically involving a single base station (BS), where the sensing transmitter and receiver are co-located and operate in full-duplex mode \cite{10465106}. In contrast, multi-static sensing leverages multiple spatially distributed receivers, offering diversity gains without necessitating full-duplex operation \cite{10494224}. Therefore, the integration of ISAC with cell-free systems gains increasing attention \cite{10516289,10742291}, in which a large number of distributed access points (APs) collaborate to provide the communication and sensing services without traditional cell boundaries \cite{10859257}. Through backhaul links, a central processing unit (CPU) connects all APs, enabling cooperative communication that enhances coverage and spectral efficiency \cite{7827017}. Cell-free ISAC addresses the key limitations of single-AP ISAC systems, such as restricted observation angles that hinder target detection, by exploiting diverse sensing observations across multiple receivers, significantly enhancing both communication and sensing performance \cite{zargari2024downlink}.

Beamforming is an essential technique that improves signal strength and quality by focusing energy in specific directions.  With perfect knowledge over the channel state information (CSI), communication and sensing performance can be optimized by designing beamforming vectors. However, in practical scenarios, acquiring perfect CSI is both computationally expensive and operationally challenging due to hardware constraints, measurement overhead, and dynamic channel conditions. To overcome these limitations, beam tracking is employed in 3GPP by periodically sweeping a predefined set of beams oriented in various directions to locate the target \cite{8458146}. It resembles an omnidirectional beampattern, transmitting equal power across all angles to facilitate target exploration throughout the entire angular domain. Nevertheless, such exhaustive beam sweeping incurs significant overhead, making it less practical for mobile target detection scenarios. As a result, authors in \cite{9091558} utilized the particle swarm optimization (PSO) technique to optimize beamforming, ensuring that a minimum signal-to-noise ratio (SNR) constraint was met while eliminating the need for CSI. Although PSO significantly reduces computational costs, it is susceptible to premature convergence, especially when the swarm gets stuck in local minima, posing threats to complex optimization problems \cite{poli2007particle}. To address these challenges, deep learning (DL)-based approaches have been widely explored \cite{9791349,9314253,10543024}. In \cite{9314253}, deep neural networks (DNN) were applied to estimate beam quality based on partial beam measurements and select the beam that maximizes the SNR. In \cite{9791349}, long short-term memory (LSTM) was employed to implicitly learn features from historical channel data and directly predict the beamforming matrix for the next time slot, aiming to maximize the average achievable sum-rate. Authors in \cite{10543024} established an end-to-end predictive beamforming framework that employs convolutional neural network (CNN)-based encoders to extract channel features and a fully connected network (FCN)-based decoder to determine the optimized beam. Nonetheless, DL-based methods typically require a large amount of labeled data for training, which is costly to collect in complex environments. Furthermore, in dynamic channel environments, the lack of a stable training dataset results in inadequate training, leading to slow convergence and insufficient generalization ability during online deployment. Consequently, deep reinforcement learning (DRL) has emerged as a promising solution.

As a model-free algorithm, DRL optimizes decisions through agent-environment interaction, rather than conducting global modeling of the environment \cite{9689948}. It does not rely on precise environmental parameters; instead, it is trained by iteratively updating an online policy according to the environment feedback. To be specific, authors in \cite{9269463} developed a DRL-based beam training method, interacting with the environment to capture dynamic spatial patterns and decide the beam search range without precise CSI. Authors in \cite{9448095} introduced a DRL framework to jointly optimize the beam selection and digital precoding matrices, subject to constraints on transmit power and the selection matrix structure. To achieve a dynamic trade-off between beam selection quality and beam sweep overhead, authors in \cite{10210616} proposed a novel action space in which actions can dynamically adjust the size of the beam sweep subset based on high- and low-frequency channel propagation laws. By formulating a cognitive multi-target detection problem, authors in \cite{9362239} optimized transmitted waveforms to ensure satisfactory detection performance under pre-determined false alarm rate constraint. Additionally, several recent studies have successfully combined ISAC and DRL to address challenges in trajectory planning and resource allocation under dynamic aerial environments \cite{11072035}. However, in many practical scenarios, generating online data through continuous environment interaction is infeasible due to the high operational cost, safety concerns, and potential risks involved \cite{levine2020offline}. Thus, the contradiction between online DRL algorithms and the practical requirements of low latency and low energy consumption poses one of the major obstacles to their widespread adoption in future 6G networks. Fortunately, offline DRL was proposed as a promising solution to address this issue \cite{10078377}. It mitigates the high costs of online interaction by utilizing only previously collected data, which typically contains sequences of state-action transitions, to learn the optimal policy without exploring or interacting with the environment during the learning process. Nevertheless, since agents rely on a fixed dataset, the distribution of states and actions within the dataset often fails to cover the full range of potential situations the agent may encounter during deployment. This results in a distribution shift \cite{kumar2019stabilizing}, where the learned policy may overestimate or underperform in unexplored regions of the state space, causing the agent to perform poorly in real-world applications. Although the conservative Q-learning (CQL) framework proposed in \cite{kumar2020conservative} prevented overly optimistic estimate and reduced policy deviation from the dataset by constraining the Q-value, its conservativeness may hinder policy improvement and limit the exploration of better actions. To enhance the diversity and realism of the offline dataset across various scenarios, we resort to digital twin (DT) technology.

Digital twin has attracted growing interest in recent years \cite{10198573,10750041,10570512,10778658,10522623,8764584}. By creating virtual replicas of real-world wireless environments, it facilitates the exploration of channel characteristics, system optimization, and the development of advanced signal processing techniques \cite{10198573}. In our prior works, a denoising diffusion probabilistic model-based DT network was proposed to address the data scarcity issue often encountered in DL \cite{10750041}. By constructing a virtual data repository, the DT demonstrated effective performance in assisting channel estimation and target detection. A similar approach was introduced in \cite{10570512}, where a conditional generative adversarial network (cGAN)-based DT was designed to improve sensing channel estimation performance. However, these efforts largely remain focus on precise modeling without incorporating dynamic policy adjustments. To enable real-time network control and continuously validate and adjust the optimal state of the physical network, authors in  \cite{10778658} and \cite{10522623} leveraged DT for tasks such as channel calibration and networks slicing. While these approaches reduce the interaction frequency with physical entities and enhance the efficiency of the learning process \cite{8764584}, the adjustments and optimization of DT network still rely on the interaction with real world. 

Motivated by the above, we propose a novel DT-assisted offline DRL approach for beam selection in a cell-free ISAC MIMO system. Through limited data observation, we construct a virtual DT environment that closely mimics the real-world distribution to facilitate offline learning, which also serves as a way of data augmentation to enrich data diversity, enabling effective beam selection without the need for extensive online interaction with actual environment. Our main contributions are as follows:

\begin{itemize}
  \item We explore the novelty of the cell-free ISAC scenario by deriving that, under the constraint of maintaining a constant false alarm rate (CFAR), the joint target detection probability across distributed APs follows a non-central chi-squared distribution. Furthermore, we formulate the beam selection task, ensuring it meets the detection performance requirements, as an optimization problem aimed at minimizing the number of beam tracking steps.
  \item We formulate the beam selection procedure as a Markov decision process (MDP) and propose a dedicated DRL framework. Distinct from prior DRL-based methods that treat the problem as a generic decision-making task, our approach incorporates domain-specific expert knowledge through reward shaping, which serves to accelerate convergence and enhance the agent's comprehension of the ISAC environment.
  \item We develop a scenario-oriented decision framework by integrating a cGAN-based DT module with offline DRL. Unlike conventional online DRL approaches that require extensive real-time interaction, our offline design significantly reduces interaction overhead while maintaining high sample efficiency. The introduced cGAN-based DT module not only enriches data diversity but also serves as a high-fidelity surrogate for the real environment, enabling robust policy learning in safety-critical or resource-constrained settings. To further improve learning stability, we integrate an additional penalty term into the loss function to mitigate overestimation problem commonly encountered in DRL. This modification directly contributes to enhanced target detection performance in complex sensing scenarios. Beyond empirical design, we provide theoretical guarantees for the convergence of the agent-DT interaction loop, and rigorously derive an upper bound on the Q-function estimation error.
  \item Numerical results highlight the effectiveness of the DT module and the superior performance of the DT-assisted offline DRL across various scenarios.
\end{itemize}

The remainder of this paper is organized as follows. Section II presents the system model and formulates the beamforming problem. Section III outlines the online DRL method as a baseline. Section IV introduces the proposed digital twin-assisted offline DRL framework and elaborates on the design of the interactive beamforming procedure. Section V provides numerical results and discussions, and Section VI concludes the paper.

The notation used in this paper follows the conventions outlined below. Boldface lowercase and uppercase letters denote column vectors and matrices, respectively. $\mathbb{C}$ represents the set of complex numbers, and $\mathcal{CN}$ refers to a circularly symmetric complex Gaussian distribution. The operators $\left( \cdot \right)^\mathrm{T}$, $\left( \cdot \right)^\mathrm{H}$, $\mathbb{E}\left( \cdot \right)$, $\|\cdot\|^2$, and $\|\cdot\|_\mathcal{H}$ indicate the transpose, conjugate transpose, expectation, squared norm of a vector or matrix, and the norm in a regenerative Hilbert space, respectively.

\section{System Model and Problem Formulation}
\begin{figure}
    \centering
    \includegraphics[width=2.2in]{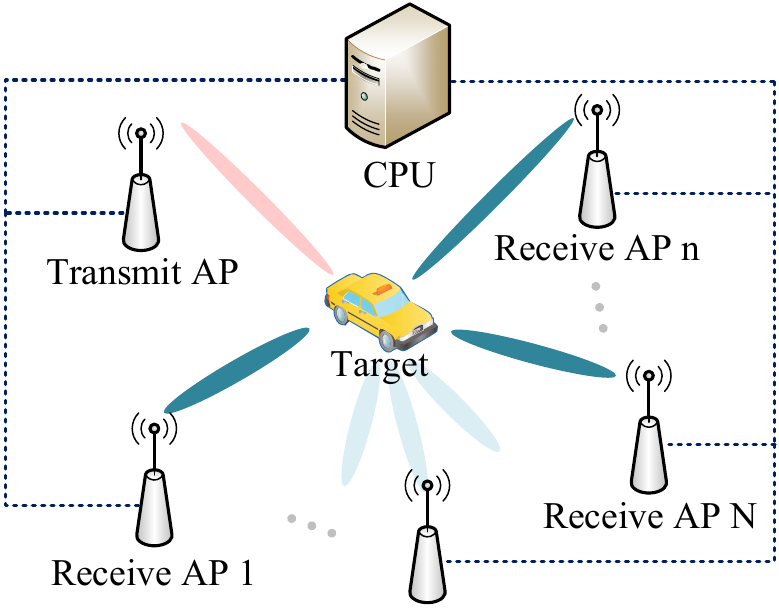}
    \caption{A cell-free ISAC system.}
    \label{System Model}
\end{figure}

Consider a cell-free massive MIMO ISAC system shown in Fig. \ref{System Model}, which consists of a transmit AP and $N$ receive APs. Each AP is equipped with an $M$-element uniform linear array (ULA), functioning as either an ISAC transmitter or receiver. In the considered configuration, all distributed APs, spatially dispersed throughout the coverage area, are interconnected to a CPU via high-speed backhaul links. These APs collaboratively detect a point-like mobile sensing target \cite{10380513, 10906066}, leveraging the coordination facilitated by the CPU. In this section, we establish the cooperative sensing model and derive the joint target detection probability under a specific beamforming design.

\subsection{Signal Model}
Let $s_{c, k}(t_s)$ denote the communication signal for the $k$-th UE and $s_r(t_s)$ denote the radar waveform for target detection at time instant $t_s \in \mathcal{T}_s$, where $\mathcal{T}_s \buildrel\textstyle.\over=[0,T_s]$ represents the duration of a block with a total duration time of $T_s$. The ISAC signal transmitted by the transmit AP is given by
\begin{equation}
\mathbf{x}(t_s)=\sum_{k=1}^K \mathbf{f}_k s_{c, k}(t_s) + \mathbf{w} s_r(t_s), \quad t_s \in \mathcal{T}_s,
\end{equation}
where $\mathbf{f}_k\in\mathbb{C}^{M\times1}$ is the beamforming vector designed for the $k$-th UE, and $\mathbf{w}\in\mathbb{C}^{M\times1}$ is the sensing beamforming vector.

Block fading channel is assumed in this work. Therefore, by neglecting the possible clutter caused by line of sight (LOS) and non line of sight (NLOS) paths due to the permanent or temporary objects, the received signal at $n$-th receive AP can be expressed as
\begin{equation}\label{sensing signal}
\mathbf{y}_n(t_s)=\alpha_{n} \sqrt{\beta_{n}} \mathbf{a}\left(\phi_n\right) \mathbf{a}^\mathrm{T}\left(\phi_0\right) \mathbf{x}(t_s-\tau_{n})+\boldsymbol{\epsilon}_n(t_s),
\end{equation}
for $n=1,\ldots,N$, where $\alpha_{n}\sim \mathcal{CN}(0,\sigma^2_{\alpha})$ is the radar cross section (RCS) of the target through the reflection path from the transmit AP to $n$-th receive AP. $\beta_{n}$ denotes the large-scale fading, which is composed of path-loss and shadow fading effect. For $n' = 0, 1, \ldots, N$, the antenna array steering vector $\mathbf{a}(\phi_{n'})\in\mathbb{C}^{M\times1}$ is given by
\begin{equation}
    \mathbf{a}(\phi_{n'})=\left[1, e^{j \pi \sin (\phi_{n'})}, \ldots, e^{j\left(M-1\right) \pi \sin (\phi_{n'})}\right]^T,
\end{equation}
with $\phi_{n'}$ being the angles of arrival or departure (AoA/AoD). $\tau_{n}=\frac{1}{c}(d_{0}+d_{n})$ is the time delay due to the target position with respect to the AP, where $c$ denotes the speed of light. $d_0$ and $d_n$ denote the distance from the transmit AP and $n$-th receive AP to the sensing target, respectively. $\boldsymbol{\epsilon}_n(t_s)\sim \mathcal{CN}(\mathbf{0},\sigma^2_{\epsilon}\mathbf{I}_M)$ is the i.i.d. zero mean circularly symmetric complex Gaussian random vector, and $\sigma_\epsilon^2$ denotes the power of the additive white Gaussian noise (AWGN).

For ease of representation, we define $\gamma_{n}\buildrel\textstyle.\over=\alpha_{n} \sqrt{\beta_{n}}$ and $\mathbf{H}_{n}\buildrel\textstyle.\over=\mathbf{a}\left(\phi_n\right) \mathbf{a}^\mathrm{T}\left(\phi_0\right)\in\mathbb{C}^{M\times M}$ as the unknown reflection coefficient and the end-to-end target response matrix from the transmit AP to the target to the $n$-th receive AP, respectively. The received echo signal in \eqref{sensing signal} can be rewritten as
\begin{equation}
    \mathbf{y}_n(t_s)\!=\!\gamma_{n}\mathbf{H}_{n}\!\left(\sum_{k=1}^K \mathbf{f}_k s_{c, k}(t_s - \tau_n) \!+\! \mathbf{w} s_r(t_s - \tau_n)\!\right)\!+\boldsymbol{\epsilon}_n(t_s).
\end{equation}
Without loss of generality, we process the received signal by a matched filter (MF) $s(t_s)$ through correlation with the known transmit waveform, where the peak of the correlation output provides an estimate $\hat{\tau}_n$ of the target delay. The MF is optimal for maximizing the output SNR with a known transmit waveform, which directly governs the detection probability. In the absence of external interference, using the MF introduces no intrinsic performance loss. For analytical tractability, we adopt the ideal assumption that the MF can perfectly tuned to the target delay \cite{6324717, 9362239}, which means $\hat{\tau}_n=\tau_n$. Consequently, we have
\begin{equation}
\begin{aligned}
\mathbf{y}_{n}= & \frac{1}{T_s} \int_{\mathcal{T}_s} \mathbf{y}_n(t_s) s_r^*(t_s-\hat{\tau}_{n}) d t_s =\gamma_{n}\mathbf{H}_{n} \mathbf{w}+\bar{\boldsymbol{\epsilon}}_{n},
\end{aligned}
\end{equation}
where $\frac{1}{T_s} \int_{\mathcal{T}_s} s_{c, k}(t_s) s_r^*(t_s)dt_s = 0$ is assumed, and $\bar{\boldsymbol{\epsilon}}_{n}=\frac{1}{T_s} \int_{\mathcal{T}_s} \boldsymbol{\epsilon}_n(t) s_r^*(t_s-\hat{\tau}_{n}) d t_s$ denotes the equivalent noise after MF processing. By aggregating all the signals $\mathbf{y}_{n}, n=1,\ldots,N$, the CPU conducts cooperative target detection based on the composite information derived from each receive AP. 

\subsection{Joint Target Detection}
Formally, joint target detection can be recast as a composite binary hypothesis test \cite{9362239,4441739}, yielding
\begin{equation}\label{eq:hypothesis}
\begin{aligned}
    \left\{\begin{array}{l}
    \mathcal{H}_0: \mathbf{y}_{n}=\bar{\boldsymbol{\epsilon}}_{n}, \quad n=1,\ldots,N,\\
    \mathcal{H}_1: \mathbf{y}_{n}=\gamma_{n}\boldsymbol{\eta}_{n}+\bar{\boldsymbol{\epsilon}}_{n}, \quad n=1,\ldots,N,
    \end{array}\right.
\end{aligned}
\end{equation}
where $\boldsymbol{\eta}_{n} \buildrel\textstyle.\over =\mathbf{H}_{n} \mathbf{w}$ is defined as the reflected signal vectors for the $n$-th receive AP.
The null hypothesis $\mathcal{H}_0$ represents the case that the sensing area contains only disturbance, which encompasses background clutter and noise, with no actual target present. Conversely, the alternative hypothesis $\mathcal{H}_1$ asserts the presence of a target within the sensing area in addition to any existing clutter and noise.

Since $\bar{\boldsymbol{\epsilon}}_{n}$ are independent of each other for different $n$, $\forall n \in \{1, \ldots, N\}$, the joint probability density function (PDF) of the spatially filtered sequence $\{\mathbf{y}_{n}\}_{n}$ under $\mathcal{H}_0$ is
\begin{equation}\label{PDF H0}
\begin{aligned}
    f_0(\{\mathbf{y}_{n}\}_{n}|\mathcal{H}_0)=\frac{1}{(\pi\sigma_\epsilon^2)^{MN}}\mathrm{exp}\left\{-\frac{\sum_{n}c_{0_{n}}}{\sigma_\epsilon^2}\right\},
\end{aligned}
\end{equation}
where
\begin{equation}\label{c0}
    c_{0_{n}}\buildrel\textstyle.\over=\|\mathbf{y}_{n}\|^2.
\end{equation}
Similarly, the joint PDF of the spatially filtered sequence $\{\mathbf{y}_{n}\}_{n}$ under $\mathcal{H}_1$ is
\begin{equation}\label{PDF H1}
\begin{aligned}
    f_1(\{\mathbf{y}_{n}\}_{n};\gamma_{n}|\mathcal{H}_1)=\frac{1}{(\pi\sigma_\epsilon^2)^{MN}}\mathrm{exp}\left\{-\frac{\sum_{n}c_{1_{n}}(\gamma_{n})}{\sigma_\epsilon^2}\right\},
\end{aligned}
\end{equation}
where
\begin{equation}\label{c1}
    c_{1_{n}}(\gamma_{n})\buildrel\textstyle.\over=\|\mathbf{y}_{n}-\gamma_{n}\boldsymbol{\eta}_{n}\|^2.
\end{equation}

Subsequently, we aim to derive the estimation over the unknown $\gamma_n$ from $f_1(\{\mathbf{y}_{n}\}_{n}; \gamma_{n}|\mathcal{H}_1)$. The negative log-likelihood function of \eqref{PDF H1} under $\mathcal{H}_1$ is proportional to
\begin{equation}
    V_1(\gamma_{n})=\mathrm{ln}(\sigma_\epsilon^2)+\frac{\sum_{n}c_{1_{n}}(\gamma_{n})}{\sigma_\epsilon^2}.
\end{equation}
Thus, minimizing $V_1$ with respect to $\gamma_{n}$ is equivalent to minimizing
\begin{equation}
\begin{aligned}
    V_2(\gamma_{n})&=\sum_{n}c_{1_{n}}(\gamma_{n})\\
    &=\sum_{n}\left(c_{0_{n}}-\gamma_{n}\bar{y}_{n}^*-\gamma_{n}^*\bar{y}_{n}+|\gamma_{n}|^2 \|\boldsymbol{\eta}_{n}\|^2\right),
\end{aligned}
\end{equation}
where $\bar{y}_{n}=\boldsymbol{\eta}_{n}^\mathrm{H}\mathbf{y}_{n}$. 
Thus, the minimization of $V_2$ with respect to $\gamma_{n}$ gives
\begin{equation}\label{hat gamma}
    \hat{\gamma}_{n}=\frac{\bar{y}_{n}}{\|\boldsymbol{\eta}_{n}\|^2}.
\end{equation}

In order to differentiate between $\mathcal{H}_0$ and $\mathcal{H}_1$, the generalized likelihood ratio test (GLRT) for the joint target detection is obtained as
\begin{equation}\label{GLRT before ln}
    \bar{\Lambda}\left(\mathbf{y}\right)=\frac{f_1(\{\mathbf{y}_{n}\}_{n};\gamma_{n}|\mathcal{H}_1)}{f_0(\{\mathbf{y}_{n}\}_{n}|\mathcal{H}_0)} \stackrel{\mathcal{H}_1}{\underset{\mathcal{H}_0}{\gtrless}} \bar{\lambda},
\end{equation}
where $\bar{\lambda}$ is the threshold and is chosen to satisfy the control of CFAR. Substituting the expressions for $f_0$ and $f_1$ formulated in \eqref{PDF H0} and \eqref{PDF H1}, and replacing the $\gamma_{n}$ with $\hat{\gamma}_{n}$ in \eqref{hat gamma}, the
decision rule can be simplified as 
\begin{equation}\label{Lambda}
\begin{aligned}
    \Lambda(\mathbf{y})& \buildrel\textstyle.\over=\ln\bar{\Lambda}(\mathbf{y})\\
    &=\frac{1}{\sigma_\epsilon^2}\left(\sum_{n} c_{0_{n}}-\sum_{n} c_{1_{n}}\left(\hat{\gamma}_{r, n}\right)\right)\\
    &=\frac{1}{\sigma_\epsilon^2}\sum_{n}\left(|\hat{\gamma}_{n}|^2 \|\boldsymbol{\eta}_{n}\|^2\right)\\
    &=\frac{1}{\sigma_\epsilon^2}\sum_{n}\left(\frac{|\bar{y}_{n}|^2}{\|\boldsymbol{\eta}_{n}\|^2}\right)\stackrel{\mathcal{H}_1}{\underset{\mathcal{H}_0}{\gtrless}} \lambda,
\end{aligned}
\end{equation}
where $\lambda\buildrel\textstyle.\over=\ln\bar\lambda$.

Under the null hypothesis $\mathcal{H}_0$ that contains only AWGN, the squared magnitude of the complex Gaussian variable $\bar{y}_{n}$ obeys
\begin{equation}
    \left|\bar{y}_{n}\right|^2\sim\sigma_\epsilon^2 \|\boldsymbol{\eta}_{n}\|^2\cdot \chi^2_2,
\end{equation}
where $\chi^2_2$ is a chi-squared random variable with $2$ degrees of freedom. Since $\Lambda(\mathbf{y})$ is the sum of $N$ chi-squared random variables, we have
\begin{equation}
	\Lambda(\mathbf{y}) \sim \chi^2_{2N},
\end{equation}
which follows a chi-squared distribution with $2N$ degrees of freedom. To ensure CFAR, we set the threshold $\lambda$ such that
\begin{equation}\label{P_FA}
\operatorname{Pr}\left\{\Lambda(\mathbf{y})>\lambda \mid \mathcal{H}_0\right\}=\int_\lambda^{\infty} p_{\Lambda \mid \mathcal{H}_0}\left(a \mid \mathcal{H}_0\right) d a \buildrel\textstyle.\over=P_{\mathrm{FA}},
\end{equation}
where $p_{\Lambda \mid \mathcal{H}_0}$ is the PDF of $\Lambda\left(\mathbf{y}\right)$ under $\mathcal{H}_0$ and
$P_{\mathrm{FA}}$ is defined as the false alarm rate. For the chi-squared distribution,
\begin{equation}
    P_{\mathrm{FA}}=1-F_{\chi^2_{2N}}\left(\lambda\right),
\end{equation}
where $F_{\chi^2_{2N}}(\cdot)$ is the cumulative distribution function (CDF) of the chi-squared distribution. Thus, the threshold $\lambda$ is given by the inverse of the CDF function.

Under the hypothesis $\mathcal{H}_1$, since the projected value obeys $\bar{y}_{n}=\hat{\gamma}_{n}\|\boldsymbol{\eta}_{n}\|^2+\boldsymbol{\eta}_{n}^\mathrm{H}\bar{\boldsymbol{\epsilon}}_{n} \sim \mathcal{CN}\left(\hat{\gamma}_{n}\|\boldsymbol{\eta}_{n}\|^2, \sigma_\epsilon^2 \|\boldsymbol{\eta}_{n}\|^2\right)$, we have
\begin{equation}
    \left|\bar{y}_{n}\right|^2\sim\sigma_\epsilon^2 \|\boldsymbol{\eta}_{n}\|^2\cdot \chi^2_2(\lambda^\mathrm{nc}_{n}),
\end{equation}
which follows a non-central chi-squared distribution, where $\lambda^\mathrm{nc}_{n}$ denotes the non-centrality parameter and is formulated as
\begin{equation}
	\lambda^\mathrm{nc}_{n}=\frac{|\hat{\gamma}_{n} \|\boldsymbol{\eta}_{n}\|^2|^2}{\sigma_\epsilon^2 \|\boldsymbol{\eta}_{n}\|^2}.
\end{equation}
In this case, $\Lambda(\mathbf{y})$ is a weighted sum of non-central chi-squared variables and obeys
\begin{equation}
    \Lambda(\mathbf{y}) \sim \sum_{n}\chi^2_2(\lambda^\mathrm{nc}_{n}),
\end{equation}
which has the following non-central chi-squared distribution properties. The sum of non-central chi-squared distributions with different non-centrality parameters $\lambda^\mathrm{nc}_{n}$ results in another non-central chi-squared distribution, i.e.,
\begin{equation}
    \Lambda(\mathbf{y})\sim \chi^2_{2N}\left(\sum\nolimits_{n}\lambda^\mathrm{nc}_{n}\right).
\end{equation}
Finally, the detection probability $P_D$ is calculated by
\begin{equation}\label{P_D}
	\operatorname{Pr}\left\{\Lambda(\mathbf{y})>\lambda \mid \mathcal{H}_1\right\}=\int_\lambda^{\infty} p_{\Lambda \mid \mathcal{H}_1}\left(a \mid \mathcal{H}_1\right) d a \buildrel\textstyle.\over=P_D,
\end{equation}
where $p_{\Lambda \mid \mathcal{H}_1}$ is the PDF of $\Lambda\left(\mathbf{y}\right)$ under $\mathcal{H}_1$.
\begin{remark}
    For a special case when $N=1$, $P_D = Q_M(\sqrt{\lambda_\mathrm{nc}},\sqrt{\lambda})$, where $Q_M(\cdot,\cdot)$ is the generalized Marcum-Q function \cite{1055327} expressing the complementary CDF (CCDF) of the non-central chi-squared distribution.
\end{remark}
\begin{remark}
While Eq.~\eqref{eq:hypothesis} assumes i.i.d. AWGN across APs, this assumption is mainly adopted for analytical tractability. In practice, the noises at spatially separated APs are independent but may exhibit distinct variances due to hardware or environmental differences. In such cases, the detection statistic becomes a sum of independent but non-identically distributed non-central chi-squared variables, whose characteristic function is
\begin{equation}
    \varphi_{\Lambda \mid \mathcal{H}_1}(t) = \prod_{n=1}^{N} (1 - 2 j \sigma_n^2\|\boldsymbol{\eta}_{n}\|^2 t)^{-1} 
    e^\frac{j \left|\hat{\gamma}_n \|\boldsymbol{\eta}_{n}\|^2 \right|^2\cdot t}{1 - 2 j \sigma_n^2 \|\boldsymbol{\eta}_{n}\|^2\cdot t},
\end{equation}
and its PDF can be obtained via inverse Fourier transform. 
This generalization does not affect the independence assumption but slightly modifies the analytical expression of the detection probability.
\end{remark}

So far,  the detection probability under the GLRT detection scheme has been derived, with the CFAR property ensured as specified in \eqref{P_FA}. Notably, for target detection at a specific position or within the same block, beamforming serves as a pivotal factor in determining the detection probability.

\subsection{Problem Formulation}
Unlike the case with perfect knowledge of $\{\mathbf{H}_n\}_{n = 1}^N$, where the detection probability $P_D$ can be maximized by directly optimizing the beamforming vector $\mathbf{w}$, this study addresses a beam selection problem based on a predefined codebook $\mathcal{W}$, without requiring accurate CSI, which would otherwise introduce substantial overhead.
Without loss of generality, we simply adopt the discrete Fourier transform (DFT)-based codebook in this work. 

Mathematically, the optimization problem is formulated to minimize the total number of beam tracking steps as follows
\begin{subequations}
\begin{align}
    (P1): \min _{T, \{\mathbf{w}_t\}_{t=1}^T} & {T} \notag\\
     \text{ s.t. } \quad & \mathbf{w}_t \in \mathcal{W}, \quad t= 1, \ldots, T,\label{constrain w}\\
     & P_{D}^t < \zeta, \quad t = 1, \ldots, T-1, \label{constrain Pd2}\\ 
     & P_{D}^T \geq \zeta,\label{constrain Pd}
\end{align}
\end{subequations}
where $T$ denotes the total number of tracking steps required in a specific scenario. The constraint in \eqref{constrain w} ensures that all the beamforming vectors are selected from the predefined DFT codebook. $\zeta$ in \eqref{constrain Pd2} and \eqref{constrain Pd} specifies a threshold set to guarantee satisfactory detection performance, which is ultimately achieved at the final tracking step $T$.
We propose our beam tracking policy by sequentially adjusting the transmitted beam codeword in response to environmental feedback. This approach obviates the need for exhaustive beam sweeping, thereby significantly reducing the communication overhead while maintaining efficient target detection. 

\textbf{Since the detection result from the previous decision directly impacts the subsequent action, rendering this process well-suited for modeling as an MDP.} To tackle this sequential decision-making problem, we utilize DRL to develop an interactive beam tracking strategy. DRL is particularly advantageous in this context due to its ability to handle high-dimensional state spaces and learn optimal policies through interactions with the environment, making it highly effective for complex beam tracking tasks.

\section{Online Deep Reinforcement Learning-Based Beam Tracking Method}\label{section_online}
In this section, we establish the online DRL framework, which is trained using an online policy that iteratively interacts with the actual environment. The beam tracking procedure is first analyzed and modeled as an MDP. Subsequently, the optimization problem $(P1)$ is addressed by systematically designing the key components of the DRL model, including the agent, the state-action transition, and the reward function. Furthermore, considering the sparsity features, reward shaping is introduced to facilitate agent learning. Finally, the network architecture and training process are illustrated.

\subsection{MDP Formulation}
MDP is defined by a 4-tuple $\{\mathcal{S},\mathcal{A},\mathcal{P},\mathcal{R}\}$, where $\mathcal{S}$ is the state space, $\mathcal{A}$ is the action space, $\mathcal{R}: \mathcal{S}\times\mathcal{A}\rightarrow \mathbb{R}$ is the immediate reward after executing action $a_t\in\mathcal{A}$ at state $\mathbf{s}_t\in\mathcal{S}$, and $\mathcal{P}$ is the state transition probability function. According to Markov property, the condition $p(\mathbf{s}_{t+1}|\mathbf{s}_t,a_t,\mathbf{s}_{t-1},a_{t-1},\ldots,\mathbf{s}_0,a_0)= p(\mathbf{s}_{t+1}|\mathbf{s}_t,a_t)$ is satisfied. During the process of beam tracking, the transmit AP acts as an agent to learn decision-making and control policies in the system. It continuously interacts with the environment by transmitting beam codewords $\mathbf{w}_t$, receiving feedback in the form of detection probability $P_D^t$, and adjusting its subsequent beam codewords $\mathbf{w}_{t+1}$. Based on these foundations, the specific design of MDP in our considered beam tracking strategy is as follows.

\subsubsection{State}
The state vector at $t$-th stage is composed of three parts: the index of beam codeword selected at the current stage $\mu_{t}$, the corresponding detection probability $P_D^{t}$, and the feedback flag $f_g^t$. To be specific, $\mu_t$ is defined as the selected column index of the codebook $\mathcal{W}$, i.e., $\mathbf{w}_t=[\mathcal{W}]_{\mu_t}$. Furthermore, an feedback flag $f_g^t$ is introduced to mark the effectiveness of the previous action, expressed as
\begin{equation}
    f_g^t = 
    \begin{cases}
        0, &t = 0,\\
        \operatorname{sgn}\left(\left({P}_D^t-{P}_D^{t-1}\right) \cdot a_t\right), &t \ge 1,
    \end{cases}
\end{equation}
where $\operatorname{sgn}\left(\cdot\right)$ is the sign function, $a_t$ is the action operated at $t$-th stage. When $f_g^t = 1$, the beam adjustment enhances the detection probability compared with the previous step; when $f_g^t = -1$, it indicates a degradation; and $f_g^t = 0$ corresponds to the initialization stage. This interpretable variable thus provides a concise and physically meaningful feedback that guides the agent to progressively enhance beam alignment and detection performance. Consequently, the state vector at $t$-th stage is mathematically defined as $\mathbf{s}_t \buildrel\textstyle.\over=[\mu_t, {P}_D^t,f_g^t]^\mathrm{T}$.

\subsubsection{Action}
The action is defined as the change in the beam codeword index $a_t=\Delta\mu_t\in\{-\tau_\mu,\ldots,-1,1,\ldots,\tau_\mu\}$. This implies that the agent selects a column by either advancing or retreating within a maximum predefined range $\{-\tau_\mu,\tau_\mu\}$.

\subsubsection{Reward}
Since our goal is to minimize the total number of tracking steps for the agent accomplishing the target sensing task, we consider the term `$-1$' as the reward to penalize the agent to finish the task as soon as possible. Maximizing the cumulative reward is equivalent to minimizing the tracking steps. Therefore, the reward function is designed as follows:
\begin{equation}
r_t = \begin{cases}
            -1, & {P}_D(t) < {\zeta}, \\
		b_0, & \text { otherwise},\end{cases}\label{reward_ori}
\end{equation}
where a positive reward $b_0 > 0$ is designed to indicate the accomplishment of the target sensing task. This achievement is a constant for any policy, thus, does not influence the optimality of the optimization problem.

\subsubsection{State Transition Probability}
$p(\mathbf{s}_{t+1}|\mathbf{s}_t,a_t)$ represents the transition probability from state $\mathbf{s}_t$ to state $\mathbf{s}_{t+1}$ after executing action $a_t$. Particularly, the transitions the beam codeword index $\mu_t \to \mu_{t+1}$ and the feedback flag $f_g^t \to f_g^{t+1}$ are deterministic. The transition of the detection probability $P_D^t \to P_D^{t+1}$ is influenced by stochastic factors, including AWGN, the estimation errors in $\{\hat{\gamma}_n\}_{n=1}^N$, and the mobility of the target.

\subsection{Reward Shaping}
\begin{figure}[t]
    \centering
    \includegraphics[width=2.5in]{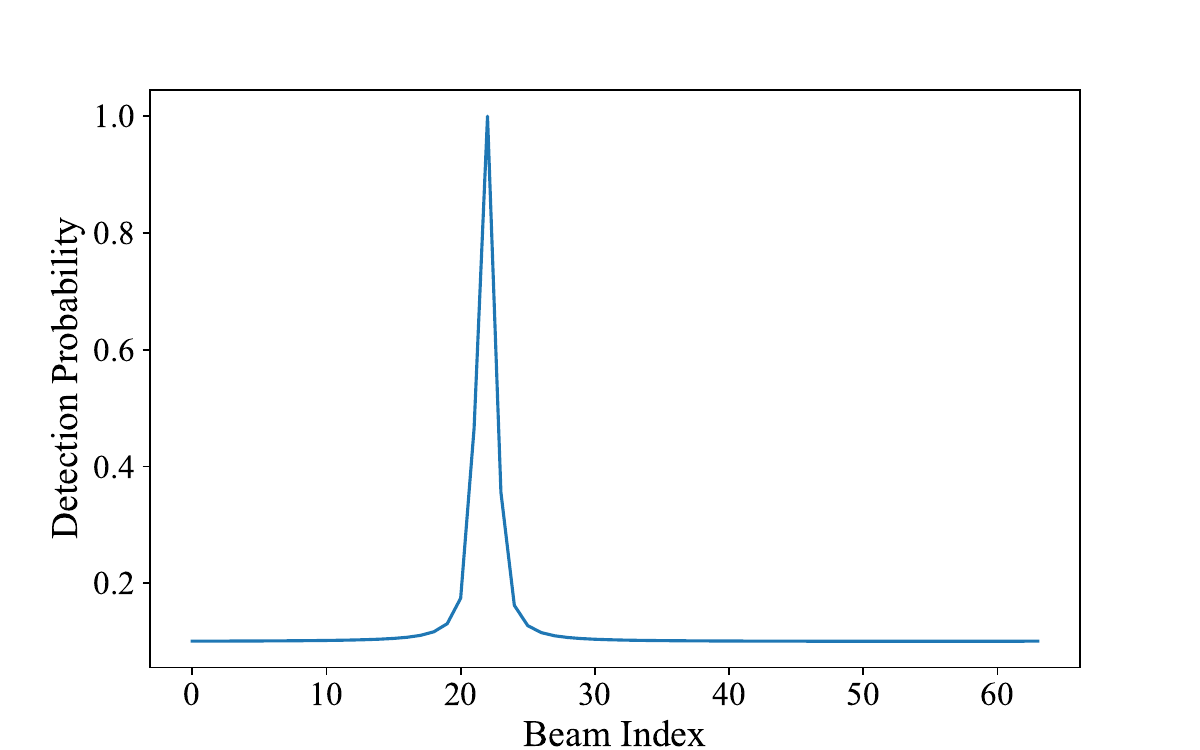}
    \caption{The CDF curve of detection probability.}
    \label{CDF of Pd}
\end{figure}
As illustrated by the CDF curve in Fig. \ref{CDF of Pd}, most beam codewords fail to detect the target, with only a few satisfying the detection performance criteria. Thus, the designed reward function in \eqref{reward_ori} closely aligns with the optimization objective, however, is severely sparse, exhibiting a flat landscape at $-1$ with a single peak at $b_0$ when the goal is achieved. Such a situation will require the agent to spend a significant amount of time on exploration.

Reward shaping \cite{ng1999policy} is defined as the addition of an extra reward signal that encodes some heuristic knowledge of the system  designer or domain expert, addressing the reward sparsity problem.

\begin{proposition}
The optimality is guaranteed if and only if the introduced shaping reward is a potential-based shaping function in terms of anteroposterior states. In this work, we define the shaping reward for transition from $\mathbf{s}_t$ to $\mathbf{s}_{t+1}$ as
\begin{equation}
    r_t^\mathrm{sh} = \rho \Phi(\mathbf{s}_{t+1}) - \Phi(\mathbf{s}_{t}),
\end{equation}
where $\rho$ is the discount factor, and the potential-based function
\begin{equation}
    \Phi(\mathbf{s}_{t}) = \frac{b_1}{1 + e^{-b_2 \cdot {P}_D(t)}}
\end{equation}
is designed only related to the current state, where $b_1$ controls the amplitude and scales the magnitude of the shaping reward, while $b_2$ determines the steepness of the sigmoid function. The sigmoid form is employed to emphasize slight variations in low values of ${P}_D(t)$. 
\end{proposition}
The reward function after applying the reward shaping can be expressed as $r_t + r_t^\mathrm{sh}$, which physically means that we encourage the agent to increase the detection probability as much as possible. Reward shaping is proposed in this paper as a way to \textbf{speed up learning and make the agent gain better knowledge of the ISAC environment}. 

Consequently, our objective is to maximize the cumulative reward across all stages, expressed as $\sum_t (r_t+r_t^\text{sh})$. While DRL lacks explicit knowledge of the mathematical relationship between the beamforming vector and the target detection probability, it effectively optimizes the beam tracking strategy through iterative interactions with the environment.

\subsection{Online DRL-Based Beam tracking}
The action-value function of the agent executing action $a_t$ at state $\mathbf{s}_t$ is defined as
\begin{equation}
    Q(\mathbf{s}_t, a_t; \theta_q) \buildrel\textstyle.\over = \mathbb{E} \left[\sum_{t'=0}^{\infty}  \rho^{t'}r_{t'}|\mathbf{s}_{0}=\mathbf{s}_{t}, a_{0}=a_t; \theta_q\right],
\end{equation}
which represents the expected long-term cumulative reward achievable under current policy characterized by the trainable parameters of the Q-network $\theta_q$. The purpose of our algorithm is to learn a policy that obtains the maximum action-value function. Subsequently, the designs of the Q-network structure and the interactive process are demonstrated as follows.

\subsubsection{Q-Network Structure}
The network design is demonstrated in Fig \ref{GAN-DQN-Network}. Specifically, for its input, to enable the column index value to serve as learnable parameters during training, we transform them into a sinusoidal embedding $\boldsymbol{\mu}_t = [\bar{\mu}_{t,1}, \ldots, \bar{\mu}_{t,I}]^\mathrm{T} \in \mathbb{R}^{I\times 1}$ using sine and cosine functions with varying frequencies, inspired by the transformer model \cite{vaswani2017attention}, yielding
\begin{equation}\label{sinusoidal}
    \bar{\mu}_{t,2i}=\sin\left(\frac{\mu_t}{1000^{\frac{2i}{I}}}\right), \quad
    \bar{\mu}_{t,2i-1}=\cos\left(\frac{\mu_t}{1000^{\frac{2i}{I}}}\right),
\end{equation}
where $I$ represents the total dimensions, allowing for a more effective representation of the index value in the neural network. By mapping these discrete indices into a continuous embedding space, the model can better capture and leverage the underlying relationships between indices during training. Additionally, to emphasize changes in $P_D^t$ when its value is relatively low, a variant sigmoid function is employed as follows
\begin{equation}
    \bar{P}_D^t = \sigma_v(P_D^t) \buildrel\textstyle.\over= \frac{\frac{1}{1 + e^{-b_3 \cdot {P}_D^t}} - \frac{1}{2}} {\frac{1}{1 + e^{-b_3}} - \frac{1}{2}}.
\end{equation}
This design guarantees a continuous mapping of values within $[0,1]$ to the same range, while significantly enhancing the sensitivity of the network to capture subtle variations in ${P}_D^t$ around $0$.

As for the output, a dueling network structure is used to split the Q-value into two components: a value function $U(\mathbf{s}_t)$ representing how good the state is, and an advantage function $A(\mathbf{s}_t, a_t)$ representing how much better an action is compared to others in that state. Its mathematical representation is  \cite{wang2016dueling} 
\begin{equation}
 Q(\mathbf{s}_t,a_t)=U(\mathbf{s}_t)+A(\mathbf{s}_t,a_t)-\max_{a} A(\mathbf{s}_t,a),
 \end{equation}
which allows the network to separately estimate the value of being in a state and the relative advantages of each action, helping improve learning efficiency, especially in states where multiple actions have similar Q-values.
\subsubsection{Interactive Process}
In this work, we employ the double DQN (DDQN) framework to mitigate the overestimation bias commonly encountered in standard value-based DRL policy when estimating Q-values, which decouples the action selection from Q-value estimation. To be exact, for a given state $\mathbf{s}_t$, the target for the Q-value is computed as
\begin{equation}
    V_t=r_{t}+\rho Q^{-}\left(\mathbf{s}_{t+1}, \arg \max _{a} Q\left(\mathbf{s}_{t+1}, a; \theta_q\right); \theta^{-}_q\right),
\end{equation}
where $Q^-(\cdot)$ is the target Q-value computed using the target network $Q^-$ with parameters $\theta^-_q$. $Q(\cdot)$ represents the Q-value function for the next state $\mathbf{s}_{t+1}$ and action $a$, computed using the current network. Consequently, the loss function in Dueling DDQN is defined as the mean squared error (MSE) between the predicted and target Q-values:
\begin{equation}\label{Q loss}
    \mathcal{L}_Q=\mathbb{E}\left[\left(V_t-Q(\mathbf{s}_t,a_t;\theta_q)\right)^2\right].
\end{equation}

\subsubsection{Practical Implementation}
During the practical deployment stage, the well-trained Q-network is directly used for real-time decision-making. The resulting policy is characterized by the optimal action-value function $Q^*(\mathbf{s},a;\theta_q^*)$, where $\theta_q^*$ denotes the optimal network parameters obtained after convergence. Given the current observation $\mathbf{s}_t$, the transmitter selects its beam index by
\begin{equation}\label{eq: DRL online deployment}
    a_t=\arg \max_a Q^*(\mathbf{s}_t,a;\theta_q^*),
\end{equation}
requiring only a single forward pass with negligible latency.


\begin{figure}[t]
    \centering
    \includegraphics[width=3.3in]{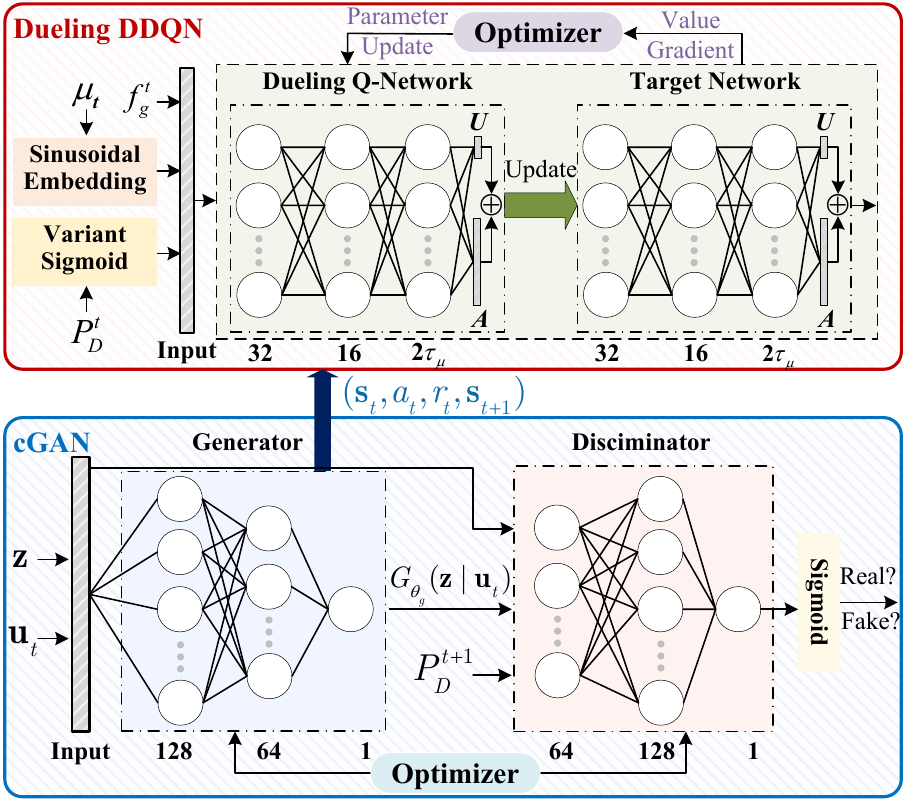}
    \caption{Network architectures of dueling DDQN and cGAN.}
    \label{GAN-DQN-Network}
\end{figure}

\section{Digital Twin-Assisted Deep Reinforcement Learning for Interactive Beamforming}
In the previous section, we have proposed the online DRL method for the target sensing task. Although online DRL has shown strong potential in addressing sequential decision-making problems, it encounters significant practical challenges in real-world applications. Specifically, it requires extensive interactions with the environment, which can be costly, time-consuming, and potentially unsafe, particularly in dynamic ISAC systems. Moreover, continuous data collection often leads to inefficiencies, as the agent often needs to explore suboptimal actions before converging to an optimal policy. These limitations underscore the need for alternative approaches that can better handle real-world constraints.

\begin{figure*}
    \centering
    \includegraphics[width=6.5in]{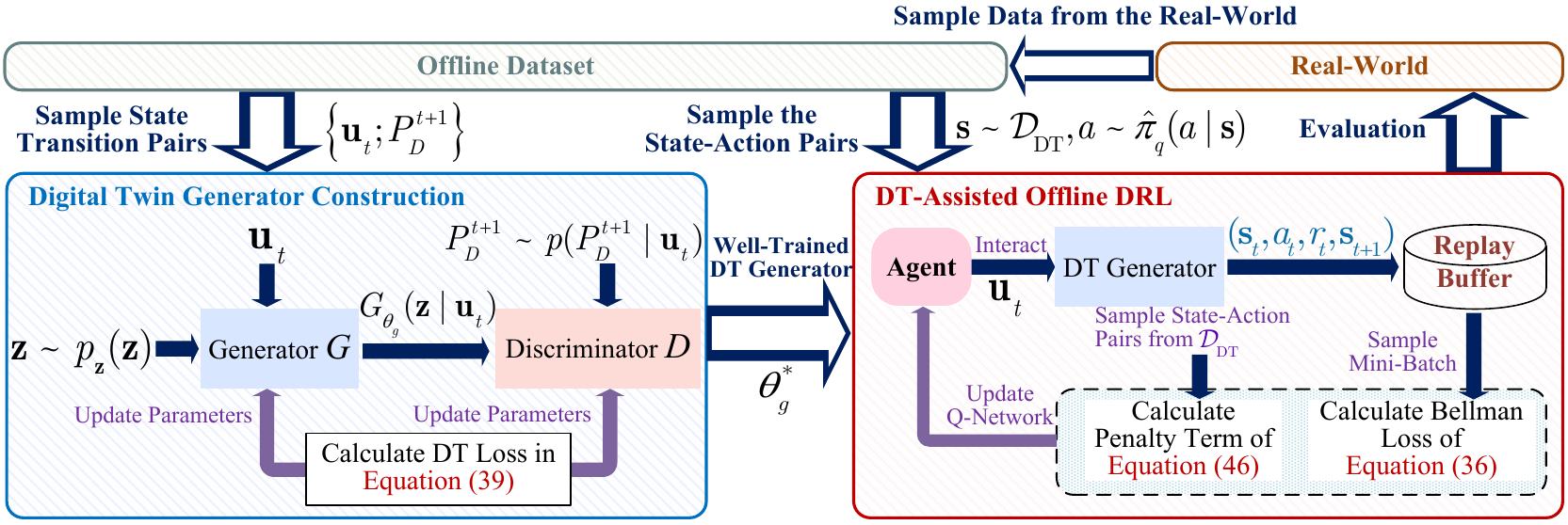}
    \caption{DT-assisted offline DRL method for beam selection.}
    \label{System Network}
\end{figure*}

To address the aforementioned issues, we propose a DT-assisted offline DRL method in this section, which comprises the DT construction phase and the DT-assisted offline DRL training phase. In the design of our proposed method, both of the two phases leverage pre-collected datasets for training, allowing the agent to learn and improve \textbf{without the need for ongoing interaction with the environment}. The specific designs of these two phases are illustrated in Fig. \ref{System Network} and analyzed below.

\subsection{The DT Construction}
\subsubsection{Problem Formulation}
To address the challenge of modeling the relationship between beam tracking strategies and detection probabilities, we define a probabilistic framework. To be specific, under a given beam selection scheme, the detection probability at the next stage follows a conditional distribution denoted as $p(P_D^{t+1}|\mathbf{u}_t)$, where $\mathbf{u}_t=\left[\mu_{t+1}, \mu_{t}, P_D^t, f_g^t\right]^\mathrm{T}$ captures both the current state of the system and the action taken at the current stage, which together influence the transition to the next state and the associated reward. 

To ensure a robust estimation of detection probabilities across varying conditions, we aim to construct a DT module, which generates an approximated distribution $\hat{p}(P_D^{t+1}|\mathbf{u}_t)$ that aligns with the true distribution $p(P_D^{t+1}|\mathbf{u}_t)$ over the candidate beamforming vectors $\mathbf{w}\in \mathcal{W}$. With the help of DT, our objective can be formalized as the minimization of a distributional distance metric
\begin{equation}
    \min_{\theta_g}\mathcal{L}_D\left({p}(P_D^{t+1}|\mathbf{u}_t),\hat{p}(P_D^{t+1}|\mathbf{u}_t; \theta_g)\right), \quad \forall \mathbf{w}\in \mathcal{W},
\end{equation}
where $\mathcal{L}_D$ quantifies the distance between the true and approximated distributions. $\hat{p}$ is the approximated distribution. $\theta_g$ denotes the parameters in the DT network to be trained. In this work, $\mathcal{L}_D$ is computed by the Jensen-Shannon (JS) divergence.

\subsubsection{Data Acquisition}
The construction of a reliable DT relies on the availability of an initial dataset to guide its training. This dataset encompasses pairs of feature vector $\mathbf{u}_t$ (i.e., the input) and the corresponding $P_D^{t+1}$ (i.e., the label) sampled from the historical state-action transitions. By learning from these pairs, the DT captures the underlying patterns in the environment and makes accurate predictions about future states.

\subsubsection{Network Construction}
To fit the distribution, our DT module employs the cGAN architecture, consisting of a generator $G$ and a discriminator $D$, both constructed using DNNs, as shown in Fig. \ref{GAN-DQN-Network}. The two networks engage in a zero-sum game: as one improves, the other must adapt to stay competitive. Specifically, during the forward pass of the generator, the conditional input $\mathbf{u}_t$ is combined with latent random noise $\mathbf{z}$ and passed through the dense layers with ReLU activations and batch normalization to generate the final detection probability output $P_D^{t+1}$. Meanwhile, the discriminator acts as a binary classifier that distinguishes between real and generated data, using ReLU activations and a final Sigmoid function to output a probability. This adversarial training process drives both networks to improve over time, formulated as
\begin{equation}\label{GAN adversarial loss}
\begin{aligned}
    &\mathcal{L}_\mathrm{DT}(\theta_d, \theta_g)=\\
    &\underset{G}{\min} \, \underset{D}{\max} \,\mathbb{E}\!\left[\log D_{\theta_d}\!\left(P_D^{t+1}\right)\!+\!\log \left(1 \!-\!D_{\theta_d}\!\left(G_{\theta_g}\!\left(\mathbf{z}|\mathbf{u}_t\right)\right)\right)\right],
\end{aligned}
\end{equation}
where $\theta_g$ and $\theta_d$ are the parameters of $G$ and $D$, respectively. The process iterates until the Nash equilibrium is achieved, at which point the generator produces data that is indistinguishable from real data to the discriminator. In this equilibrium, the discriminator outputs a probability of 0.5 for all inputs, indicating its inability to distinguish between real and fake data. According to \cite{goodfellow2020generative,10669210}, such a balance is equivalent to minimizing the JS divergence between the approximated distribution $\hat{p}(P_D^{t+1}|\mathbf{u}_t)$ and true distribution $p(P_D^{t+1}|\mathbf{u}_t)$. Detailed hyperparameters are provided in Section \ref{Simulation Setup}.

After being well-trained, the generator $G$, with well-trained parameters $\theta_g^*$, serves as a DT tool, mapping beam selections at the transmit AP to target detection probability obtained at CPU. The DT also captures the distribution of detection probabilities across various beam selection strategies and dynamic target positions, enabling high-precision pre-validation through data augmentation. \textbf{Notably, with the DT module, a virtual data repository is created, enriching data diversity via latent inputs and providing data to replay buffer. }As a result, the DRL agent no longer requires direct interaction with the physical environment but instead interacts with the DT, obtaining highly probable detection probabilities and significantly reducing interaction overhead.

Subsequently, we provide the theoretical analysis of convergence properties of our method. For easy of representation, we define the contraction mappings of the Bellman operators in the real world and DT as $\mathcal{T}_\mathrm{RW}\left(\cdot\right)$ and $\mathcal{T}_\mathrm{DT}\left(\cdot\right)$, respectively. Specifically, the state transition probabilities in the real world and the DT are denoted as $p_\mathrm{RW}(\mathbf{s}'|\mathbf{s}, a)$ and $p_\mathrm{DT}(\mathbf{s}'|\mathbf{s}, a)$, respectively, while the corresponding reward functions are represented as $r_\mathrm{RW}(\mathbf{s}, a, \mathbf{s}')$ and $r_\mathrm{DT}(\mathbf{s}, a, \mathbf{s}')$. Furthermore, we define the Q-error function at the $e$-th iteration step as 
\begin{equation}
    \Delta_e(\mathbf{s}, a) \buildrel\textstyle.\over= Q_e(\mathbf{s}, a) - Q^*(\mathbf{s}, a),
\end{equation}
where $Q_e(\mathbf{s}, a)$ denotes the Q-value at the $e$-th iteration, and $Q^*(\mathbf{s}, a)$ denotes the optimal Q-value. Based on this formulation, we derive the following theorem.
\begin{theorem}\label{theorem 1}
    As iteration proceeding, the Q-error function satisfies
    \begin{equation}
        \|\Delta^*(\mathbf{s}, a)\|\buildrel\textstyle.\over = \lim_{e\to\infty} \|\Delta_{e}(\mathbf{s}, a)\| \le \frac{\xi_{\mathrm{max}} \delta_\mathcal{T}}{1-\rho},
    \end{equation}
    if the following conditions hold:
    \begin{itemize}
        \item Learning rate $\sum_{e=1}^\infty \xi_e = \infty$ and $\sum_{e=1}^\infty \xi_e^2 < \infty$ uniformly w.p. 1, and $\xi_\mathrm{max} = \max_e\xi_e$.
        \item The variance of the reward function $r_t$ is bounded.
        \item $\delta_\mathcal{T} \buildrel\textstyle.\over = \sup_{\{\mathbf{s}, a\}} \|\mathcal{T}_\mathrm{DT}\left(Q^*(\mathbf{s}, a)\right) - \mathcal{T}_\mathrm{RW}\left(Q^*(\mathbf{s}, a)\right)\|$. 
    \end{itemize}
\end{theorem}

\begin{figure*}[t]
\begin{equation}\label{delta_e+1}
    \Delta_{e+1}(\mathbf{s}, a) = (1-\xi_t) \Delta_e(\mathbf{s}, a) + \xi_t \Big(\underbrace{\sum\nolimits_{\mathbf{s}'} p_\mathrm{DT}(\mathbf{s}'|\mathbf{s}, a) \left(r_\mathrm{DT}(\mathbf{s}, a, \mathbf{s}') + \rho \sum\nolimits_{a'} \pi(a'|\mathbf{s}')Q_e(\mathbf{s}', a') \right) - Q^*(\mathbf{s}, a)}_{F_e}\Big).
\end{equation}
\vspace{-0.2cm}
\begin{align}\label{F_e}
    F_e \!=\! & \sum\nolimits_{\mathbf{s}'} p_\mathrm{DT}(\mathbf{s}'|\mathbf{s}, a) \left(r_\mathrm{DT}(\mathbf{s}, a, \mathbf{s}') \!\!+\!\! \rho \sum\nolimits_{a'} \pi(a'|\mathbf{s}')Q_e(\mathbf{s}', a') \right) \!\!-\!\! \sum\nolimits_{\mathbf{s}'} p_\mathrm{RW}(\mathbf{s}'|\mathbf{s}, a) \left(r_\mathrm{RW}(\mathbf{s}, a, \mathbf{s}') \!\!+\!\! \rho \sum\nolimits_{a'} \pi(a'|\mathbf{s}')Q^*(\mathbf{s}', a') \right) \notag\\
        = & \sum\nolimits_{\mathbf{s}'} p_\mathrm{DT}(\mathbf{s}'|\mathbf{s}, a) \rho \sum\nolimits_{a'} \pi(a'|\mathbf{s}')\Delta_e(\mathbf{s}', a') 
        + \sum\nolimits_{\mathbf{s}'} p_\mathrm{DT}(\mathbf{s}'|\mathbf{s}, a) \left(r_\mathrm{DT}(\mathbf{s}, a, \mathbf{s}') + \rho \sum\nolimits_{a'} \pi(a'|\mathbf{s}')Q^*(\mathbf{s}', a') \right) \notag \\
        & - \sum\nolimits_{\mathbf{s}'} p_\mathrm{RW}(\mathbf{s}'|\mathbf{s}, a) \left(r_\mathrm{RW}(\mathbf{s}, a, \mathbf{s}') + \rho \sum\nolimits_{a'} \pi(a'|\mathbf{s}')Q^*(\mathbf{s}', a') \right) \notag\\
        = & \mathcal{T}_\mathrm{DT}\left(\Delta_e(\mathbf{s}, a)\right) + \mathcal{T}_\mathrm{DT}\left(Q^*(\mathbf{s}, a)\right) - \mathcal{T}_\mathrm{RW}\left(Q^*(\mathbf{s}, a)\right).
\end{align}
\hrulefill
\end{figure*}
\begin{proof}
    Derived from the training loss, we can derive the error iterative function \eqref{delta_e+1} shown at the top of the page. According to the Bellman operator, the second term in \eqref{delta_e+1} can be rewritten as \eqref{F_e}. Therefore, according to \cite{singh2000convergence}, we have the following inequality when $e\to\infty$:
\begin{equation}
    \|\Delta^*(\mathbf{s}, a)\| \le \lim_{e\to\infty} \sum_{e'=0}^{e-1} \rho^{e-e'-1}\xi_{e'} \le \frac{\xi_{\mathrm{max}} \delta_\mathcal{T}}{1-\rho}.
\end{equation}
Specifically, if $Q^*(\mathbf{s}, a)$ is the fixed point of the DT operator $\mathcal{T}_\mathrm{DT}\left(\cdot\right)$, i.e., $\mathcal{T}_\mathrm{DT}\left(Q^*(\mathbf{s}, a)\right) = Q^*(\mathbf{s}, a)$, the error converges to zero.
\end{proof}

From a physical perspective, Theorem \ref{theorem 1} implies that the proposed learning scheme achieves stable beam selection performance even under modeling discrepancies between the DT and the real environment, as the Q-value deviation remains bounded within a finite margin during convergence.

\subsection{The DT-assisted offline DRL training}
Building upon the constructed DT, the training agent enables itself to interact and learn within the virtual environment of the DT rather than the physical real-world system. Similar to the training process mentioned in Section \ref{section_online}, when the agent issues an action command, the DT acts as a substitute of the real-world system, providing feedback on detection probabilities that are highly likely to occur in the real-world system.

More importantly, one of the main concerns for introducing DT in offline DRL method is the risk of overestimating Q-values, particularly for actions or state-action transition pairs that have not been sufficiently explored in the original dataset $\mathcal{D}_\mathrm{DT}$. Compared to the traditional offline DRL methods, the integration of a DT module offers a critical solution to the overestimation problem by enabling the generation of diverse synthetic data for underrepresented or out-of-distribution state-action pairs. Additionally, inspired by \cite{kumar2020conservative}, we introduce a conservative offline strategy which modifies the loss function by adding an additional conservative penalty term weighted by the penalty coefficient $\alpha_\mathrm{con}$, expressed as
\begin{equation}
    \mathcal{L}_\mathrm{offline} = \mathcal{L}_Q + \alpha_\mathrm{con} \mathcal{L}_\mathrm{con},
\end{equation}
where the conservative penalty term is given by
\begin{equation}
\begin{aligned}
\mathcal{L}_\mathrm{con} = & \mathbb{E}_{\mathbf{s} \sim \mathcal{D}_\mathrm{DT}} \left[ \log \sum\nolimits_{a} \exp(Q(\mathbf{s}, a; \theta_q))\right] \\
& - \mathbb{E}_{\mathbf{s} \sim \mathcal{D}_\mathrm{DT}, a \sim \hat{\pi}_q (a|\mathbf{s})} \left[ Q(\mathbf{s}, a; \theta_q) \right],
\end{aligned}
\end{equation}
where $\hat{\pi}_q$ denotes the empirical behavior policy that produce the dataset $\mathcal{D}_\mathrm{DT}$. This introduced penalty term encourages the learned Q-function to assign lower values to actions that are less frequent in the dataset and aims to \textbf{query the Q-function value at unobserved states distributions, mitigating the out-of-distribution problem}. Together, conservative penalty and the DT framework create a robust foundation for offline DRL training, ensuring that the agent's decisions are both conservative and grounded in a comprehensive representation of the environment, ultimately leading to improved and more reliable policies. The process of DT-assisted offline DRL for beam selection is summarized in Algorithm \ref{algorithm1}.

Once the offline training is completed, the DT is no longer required during practical implementation. The trained policy, as formulated in \eqref{eq: DRL online deployment}, can then be directly deployed.

\begin{algorithm}[t]
\caption{Digital Twin-Assisted Offline DRL for Beam Selection}
\label{algorithm1}	
\textbf{Initialize:} Initialize action-value function $Q$ with random weights $\theta_q$. Initialize target action-value function $Q^-$ with weights $\theta_q^- = \theta_q$. Initialize the cGAN model with random parameters $\theta_g$ and $\theta_d$.\\
\textbf{Data Collection:} Randomly select actions and interact with the environment with policy $\hat{\pi}_q$ to collect some data $(\mathbf{s}_t,a_t,r_t,\mathbf{s}_{t+1})$ and obtain a dataset $\mathcal{D}_\mathrm{DT}$.\\
\textbf{Train the Digital Twin:} Train the cGAN model $\theta_g$ and $\theta_d$ using the loss function \eqref{GAN adversarial loss} and dataset $\mathcal{D}_\mathrm{DT}$.\\
    \For{$\bar{e}=1,\ldots,\bar{E}$} 
	{
        \For{$\hat{e}=1,\ldots,\hat{E}$}
        {	
	Update the discriminator parameters $\theta_d$ via gradient ascent on:
    $ \max_{\theta_d}\mathbb{E}[\log D_{\theta_d}(P_D^{t+1})+\log(1-D_{\theta_d}(G_{\theta_g}(\mathbf{z}|\mathbf{u}_t)))].$
	}		
        Update the generator parameters $\theta_g$ via gradient descent with well-trained $\theta_d^*$:
        $\min _{\theta_g} \mathbb{E}[\log(1-D_{\theta_d^*}(G_{\theta_g}(\mathbf{z}|\mathbf{u}_t)))].$
	}
\textbf{Train the DRL Agent:}\\
    \For{$e=1,\ldots,E$} 
    {Initialize observation state $\mathbf{s}_1$.\\
    \For{$t=1,\ldots,T$}
    {Observe the current state $\mathbf{s}_t$. \\
    With probability $P_e=0.2(1-\frac{e}{E})$ select a random action $a_t$. Otherwise select $a_t=\text{argmax}_a Q(\mathbf{s}_t,a;\theta)$.\\
    Execute	beamforming action $a_t$ in the digital twin environment, interact with the well-trained generator $G_{\theta_g^*}$ and obtain the reward $r_t$ along with the next state $\mathbf{s}_{t+1}$ according to the output of it.\\
    Store transition $(\mathbf{s}_t,a_t,r_t,\mathbf{s}_{t+1})$ in the replay buffer $\Omega$.\\
    Randomly sample minibatch of transitions from $\Omega$ and train the Q-network by performing a gradient descent step with loss function \eqref{Q loss}.\\
    Update the Q-network $Q$.\\
    Every $C$ steps update the target network $Q^-$ with $\theta_q^-=\theta_q$.	
    }
    }
\end{algorithm}

\subsection{Computational Complexity Analysis}
The computational complexity of the cGAN-based DT mainly arises from the fully connected (FC) layers in both the generator and the discriminator. Let $C_{in}^{(l)}$ and $C_{out}^{(l)}$ denote the input and output dimensions of the $l$-th FC layer, respectively, and let $L_G$ and $L_D$ represent the total number of layers in the generator and discriminator. The computational complexity of a forward pass can thus be expressed as $\mathcal{O}\left(\sum_{l=1}^{L_G} C_{in,G}^{(l)} C_{out,G}^{(l)} + \sum_{l=1}^{L_D} C_{in,D}^{(l)} C_{out,D}^{(l)}\right)$.
According to the network architecture shown in Fig.~\ref{GAN-DQN-Network}, we have $L_G=3,\,C_{in,G}^{(1)}=d_{in,G},\, C_{out,G}^{(1)}=C_{in,G}^{(2)}=128,\,
    C_{out,G}^{(2)}=C_{in,G}^{(3)}=64,\, C_{out,G}^{(3)}=1;\,L_D=3,\,
    C_{in,D}^{(1)}=d_{in,D},\, C_{out,D}^{(1)}=C_{in,D}^{(2)}=64,\,
    C_{out,D}^{(2)}=C_{in,D}^{(3)}=128,\, C_{out,D}^{(3)}=1$, where $d_{in,G}$ and $d_{in,D}$ denote the input dimensions of generator and discriminator, respectively. Since the proposed network employs lightweight FC layers with moderate hidden dimensions, the overall computational cost grows linearly with the input dimension. Moreover, the computational overhead of the DT only exists in the offline training phase. Once the twin environment is constructed, no additional computational burden is introduced during actual engineering implementation.

The computational complexity of the DRL module is evaluated in a similar way. During both training episodes and online inference, the dominant computational cost arises from matrix multiplications within the neural network layers. Let $L_Q$ represent the total number of layers in the dueling Q-network, the computational complexity can be expressed as $\mathcal{O}\left(\sum_{l=1}^{L_Q} C_{in,Q}^{(l)} C_{out,Q}^{(l)}\right)$. According to the network structure illustrated in Fig.~\ref{GAN-DQN-Network}, $L_Q=3,\, C_{in,Q}^{(1)}=d_{in,Q},\, C_{out,Q}^{(1)}=C_{in,Q}^{(2)}=32, \,
    C_{out,Q}^{(2)}=C_{in,Q}^{(3)}=16, \, C_{out,Q}^{(3)}=2\tau_\mu,$ where $d_{in,Q}$ represents the input dimension.


\section{Numerical Results}
This section investigates the performance of the proposed DT-assisted offline DRL for beam selection. First, we evaluate the DT module by calculating metrics that quantify the distributional differences. Then, we compare the performance of our method with other approaches (e.g., online DRL, beam sweeping and heuristic algorithm) under various transmission conditions, CFAR constraints, and target velocities to highlight the effectiveness and superiority of our method.

\subsection{Simulation Setup}\label{Simulation Setup}
\begin{table*}[t]
	\caption{Parameters for DT and DRL training.}
        \label{table:Parameters for DT and DRL training.}
	\centering
	\begin{tabular}{c|c c c c c c} \hline		
	\bf{\tabincell{c}{Task}}& Batchsize & Epoch / Iteration & Optimizer & Learning Rate & Network Depth & Others \\ \hline
	\bf{\tabincell{c}{DT Training}} & 64 & $\bar{E}=100,\hat{E}=1$ & Adam (0.5, 0.999) & $2\times 10^{-4}$ & $L_G, L_D=3$ & -  \\ \hline
	\bf{\tabincell{c}{DRL Training}} & 16 &  $E=1000,T=64,C=200$ & Adam (0.9, 0.999)  & $10^{-4}\cdot 0.9^e$ & $L_Q=3$ & $\alpha_\mathrm{con}=0.1, |\Omega|=5\times10^4$ \\ \hline 
	\end{tabular}
\end{table*}
We consider a cell-free ISAC scenario with a transmit AP and $N=4$ receive APs distributed in the area of size 400m $\times$ 400m. To be exact, they are positioned at coordinates (250m, 250m), (100m, 100m), (100m, 400m), (400m, 100m), and (400m, 400m), respectively, ensuring equidistant placement. Each AP is equipped with $M=64$ antennas. The target trajectory follows a piecewise-linear motion model, with the target starting from a random initial position uniformly distributed within the APs' coverage area and moves randomly at a velocity of $v=0\sim5 \text{m/s}$. The beamforming codebook uniformly quantizes the azimuth range of $[0, 2\pi)$ into 64 discrete beams.
The transmitting power, the CFAR constraint $P_\mathrm{FA}$, and the detection probability threshold $\zeta$ are set to 20 dBmW, $10^{-3}$, and 0.9, respectively, unless explicitly stated otherwise.

The large-scale fading coefficient $\beta_{ij}$ between objects $i$ and $j$ is modeled to capture the effects of path loss and is formulated as: $
\beta_{ij}\mathrm{[dB]} = -34.5 - \alpha \cdot 10 \log{10} \left(\frac{d_{ij}}{d_0}\right)$,
where $d_{ij}$ denotes the Euclidean distance between objects $i$ and $j$, $d_0 = 1\text{m}$ is the reference distance, and the path loss exponent is set to $\alpha = 3.7$. 
Moreover, the power of the AWGN is computed  as $\sigma^2\mathrm{[dB]} = -174 + n_f + 10 \log_{10} \left(B\right)$, where $n_f = 9\text{dB}$ is the noise figure, and $B = 20\text{MHz}$ represents the signal bandwidth.

The pre-collected dataset used for training DT is obtained through initial agent-environment interactions by 200 episodes. The parameters for MDP modeling are listed as follows: $\tau_\mu=5, b_0=5, b_1=1, b_2=b_3=5, \rho=0.99$. Other parameters for DT and DRL training are summarized in Tabel~\ref{table:Parameters for DT and DRL training.}, where $|\Omega|$ denotes the capacity ceiling of the replay buffer.



\subsection{Fitting Accuracy of Digital Twin Module}\label{subsection: Fitting Accuracy}
Before presenting the performance of our proposed DT-assisted DRL method for beamforming, we first assess the fitting accuracy of the cGAN-based DT module in replicating the real environment, which forms the foundation for subsequent DRL training and interaction. We adopt three widely recognized metrics to evaluate the distribution differences: Kullback-Leibler (KL) divergence, Wasserstein distance, and maximum mean discrepancy (MMD), expressed as
\begin{align}
    &\mathcal{L}_{D_\mathrm{KL}}= \mathbb{E}_{\hat{p}}[\log(\frac{\hat{p}(x)}{p(x)})],\\
    &\mathcal{L}_{D_W}= \inf _{\pi_w \sim \prod(\hat{p}, p)} \mathbb{E}_{x, y \sim \pi_w}[\|x-y\|],\\
    &\mathcal{L}_{D_\mathrm{MMD}}= \sup _{\|\kappa\|_{\mathcal{H}} \leq 1} \mathbb{E}_{\hat{p}}[\kappa(x)]-\mathbb{E}_{p}[\kappa(y)],
\end{align}
respectively, where $\hat{p}$ and $p$ represent the generated data distribution and the real data distribution, respectively. $\Pi$ denotes the set of joint distributions. $\kappa$ is the kernel function, chosen as a Gaussian kernel in this work. We measure the distribution differences between detection probabilities generated by the cGAN-based DT module under the action space conditions and those obtained from the agent's interaction with the actual environment. 

\begin{figure}[t]
	\centering
	\subfloat[Generator]{
		\includegraphics[width=1.70in, page=1]{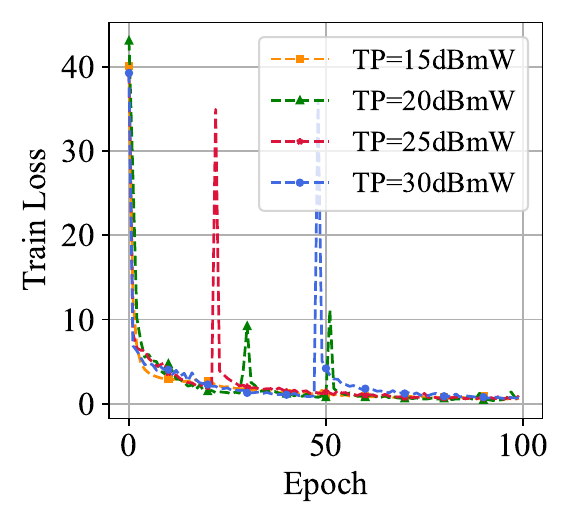}
		\label{test_loss_G}}
	\subfloat[Discriminator]{
		\includegraphics[width=1.70in, page=1]{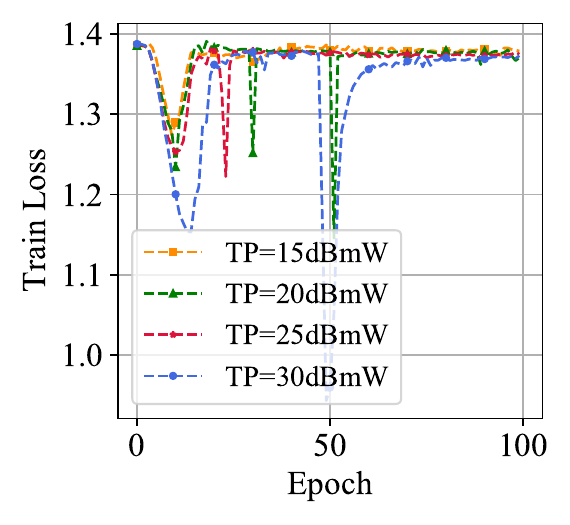}
		\label{test_loss_D}}
	\caption{Convergence behaviors of the cGAN-based DT module.}
	\label{loss curve}
\end{figure}

As shown in Fig. \ref{loss curve}, the generator and discriminator converge to their theoretical convergency values of 0.693 and 1.386 \cite{goodfellow2020generative}, respectively, within 100 epochs. This convergence, observed under transmitting power levels ranging from 15 dBmW to 30 dBmW, demonstrates the stability of the DT module. The abrupt loss fluctuation observed around epoch 50, particularly for TP = 30 dBmW, arises from the intrinsic adversarial dynamics of conditional GAN training. In such a minimax setting, temporary imbalances between the discriminator and generator can lead to sudden spikes or drops in the loss curves. This phenomenon is more evident under high-TP conditions, where stronger signals make the data distribution initially easier for the discriminator to classify, causing a temporary dominance that increases the generator loss. As training progresses, the generator adapts to restore equilibrium, producing the observed oscillatory behavior.

Additionally, Table \ref{Fitting Accuracy of DT} presents the values of three metrics across different transmitting power levels. All values remain below 0.02, indicating that the data distribution generated by the digital twin closely approximates the real one. This precise modeling lays a strong foundation for the following DRL assistance. According to the derivation in Theorem \ref{theorem 1}, the generator's error directly affects the discrepancy between $\mathcal{T}_\mathrm{DT}(Q^*)$ and $\mathcal{T}_\mathrm{RW}(Q^*)$. Consequently, a smaller generator error leads to a smaller upper bound of Q-error, enabling our proposed method to approach the optimal value.
\begin{table}
	\caption{Distribution differences between DT generated data and real data.}
        \label{Fitting Accuracy of DT}
	\centering
	\begin{tabular}{c|c c c c} \hline		
	\bf{\tabincell{c}{Transmitting Power (dBmW)}}& 15 & 20 & 25 & 30 \\ \hline
	\bf{\tabincell{c}{KL Divergence}} & 0.0139 & 0.0165 & 0.0117 & 0.0183 \\ \hline
	\bf{\tabincell{c}{Wasserstein Distance}} & 0.0162 & 0.0193 & 0.0129 & 0.0176 \\ \hline 	
	\bf{\tabincell{c}{Maximum Mean Discrepancy}} & 0.0019 & 0.0123 & 0.0078 & 0.0114 \\ \hline	
	\end{tabular}
\end{table}

\subsection{Performance Comparisons with Other Methods}
In this subsection, we explore and compare several beamforming strategies in our simulations, as detailed below:
\begin{itemize}
    \item Traditional exhaustive beam sweeping.
    \item PSO algorithm, where four particles are defined, corresponding to four steps of exploration, with the initial particle swarm randomly distributed within the codebook index range. The position of each particle represents the current beamforming vector index, while its velocity indicates the chosen tracking direction. For each particle, we calculate the detection probability of the corresponding beamforming vector in the environment.
    \item Online DRL, which interacts directly with the environment and adjusts strategies based on detection probability feedback in each episode.
    \item Offline DRL, which relies exclusively on the initially established data repository, with performance curves plotted by applying the model parameters trained in each episode to the real environment without feedback or interaction.
    \item \textbf{Our proposed DT-assisted offline DRL (DT-DRL)}, both with and without CQL.
\end{itemize}

\begin{figure*}[htbp]
\centering
    \includegraphics[height=2.5in, page=1]{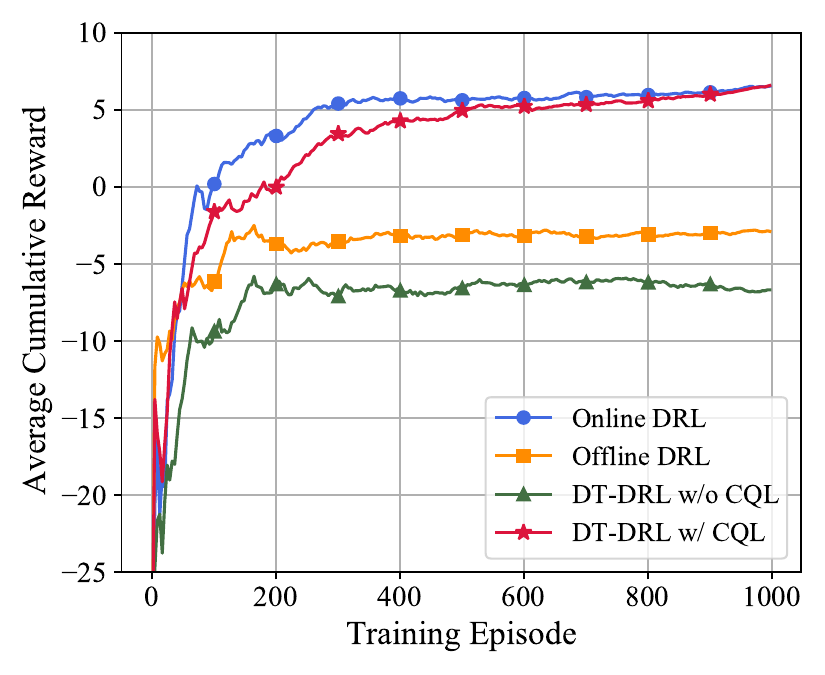}
    \label{vs other methods-episode}
    \hspace{1.2cm}
    \includegraphics[height=2.5in, page=1]{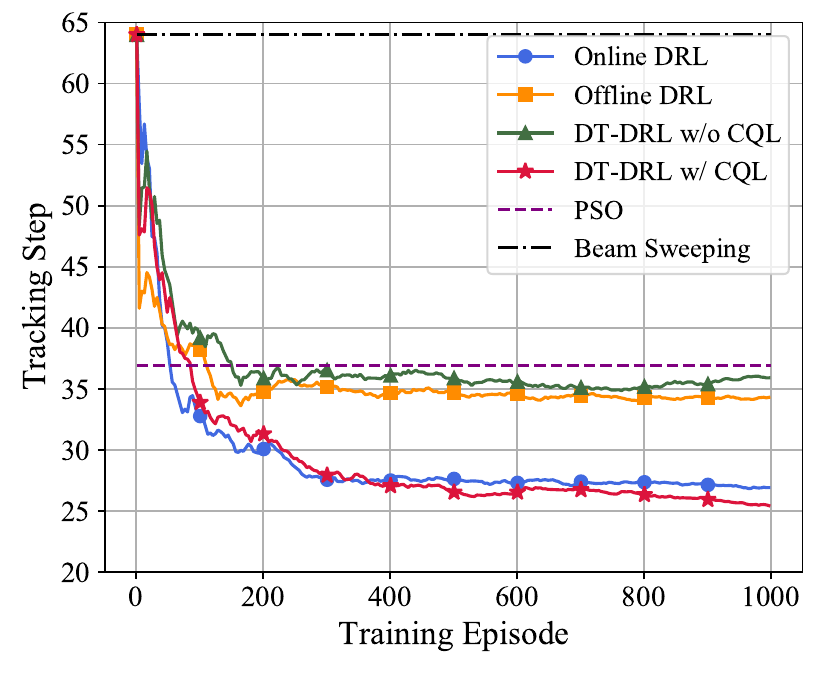}
    \label{vs other methods-step}
    \caption{Comparisons of average cumulative rewards and tracking steps between our DT-assisted DRL and other methods.}
    \label{vs other methods}
\end{figure*}

Fig. \ref{vs other methods} demonstrates the average cumulative reward and number of tracking steps required to find the optimal beamforming vector. First, it is evident that all DRL-based methods converge within 1000 episodes. Next, the declining trend of the convergence curve for all reinforcement learning algorithms closely matches the rising trend of the reward curve, thereby validating the equivalence of the designed reward. Then, DRL-based methods performs better than traditional methods due to their efficient exploration and exploitation, avoiding the local minima that PSO often encounters. 

Moreover, by comparing `Offline DRL' and `DT-DRL w/ CQL', it can be seen that offline DRL assisted by the DT module achieves significant gains compared to its counterpart without digital twins. This is because the DT module provides additional generated data for interaction and policy adjustment, enriching the data diversity and enabling the agent to overcome the constraints of limited historical data. Furthermore, DT-assisted DRL without CQL function performs the worst among the DRL algorithms due to the harmful effects of distributional shift and overestimation during exploration. Thus, the DT module and the CQL function play a crucial role in enhancing the performance of intelligent beamforming. Notably, in the early stages of training, our proposed DT-assisted DRL underperforms compared to online DRL because it relies on interaction with the DT environment, which initially cannot fully replicate the real environment or catch up real-time adaptation. By around 360-th episode, its performance surpasses that of online DRL, representing a significant improvement and demonstrating that \textbf{our proposed algorithm can achieve comparable results to online DRL while reducing interaction overhead by 80\%}.

\begin{figure}[t]
	\centering
	\includegraphics[height=1.9in, page=1]{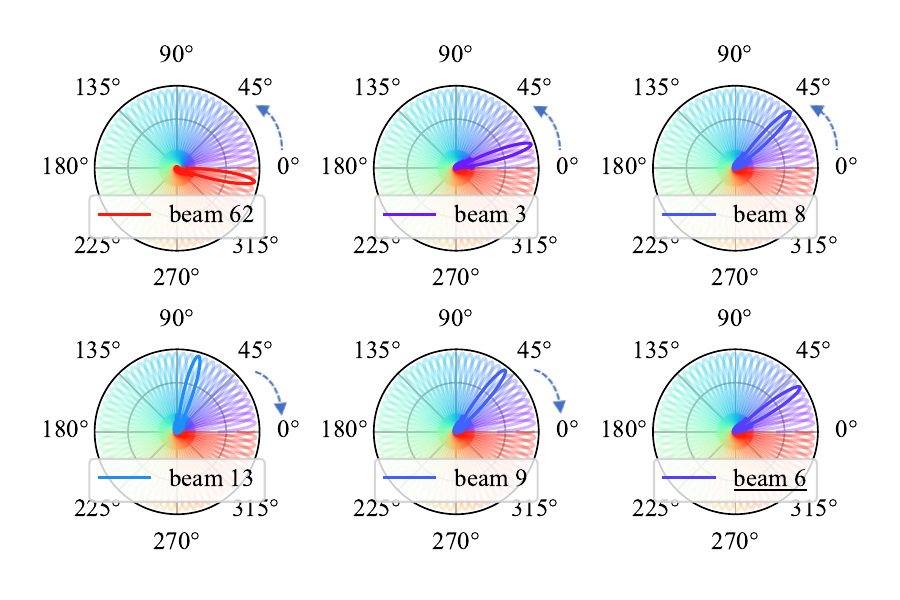}
	\caption{Beam selection using our DT-assisted DRL with CQL. Left-top: Beam 62 is selected at 1-st step with $P_D=0.1003$. Middle-top: Beam 3 is selected at 2-nd step with $P_D=0.1072$. Right-top: Beam 8 is selected at 3-rd step with $P_D=0.2136$. Left-bottom: Beam 13 is selected at 4-th step with $P_D=0.1133$. Middle-bottom: Beam 9 is selected at 5-th step with $P_D=0.2021$. Right-bottom: Beam 6 at bottom layer is selected at 6-th step with $P_D=0.9998$.}
    \label{beam plot}
\end{figure}

An illustrative example of sequential beam selection process for our proposed DT-assisted offline DRL method is shown in Fig. \ref{beam plot}. Initially, it tracks by advancing 5 beams at a time, with the detection probability increasing accordingly. When the detection probability decreases at step 4, it retreats 4 beams to explore further. At this stage, the detection probability rebounds but stays below the threshold. It then retreats 3 beams, ultimately identifying the optimal beam. This stepwise adaptation highlights the capability of our method to refine beam selection based on the virtual environment feedback to achieve satisfactory detection performance. 

Furthermore, we perform additional numerical simulations to investigate the impact of the parameter $b_2$ on the agent's learning behavior. We observe that a moderate value of $b_2$ enhances the information-capturing ability of the shaping function, enabling the agent to extract meaningful feedback in the early stage while still pursuing improved performance in later stages. In contrast, an excessively small $b_2$ weaken the reward shaping effect, preventing the agent from learning effectively. Conversely, an excessively large $b_2$ causes the reward to saturate too quickly, leading the agent to prematurely settle for suboptimal detection probabilities rather than continuing to explore better strategies. As a result, both extremes of $b_2$ severely degrade the overall convergence and final performance, whereas moderate values strike the desired balance between exploration and stability.

\subsection{Performance under Different Levels of Transmitting Power}
\begin{figure*}[htbp]
\centering
	\includegraphics[height=2.5in, page=1]{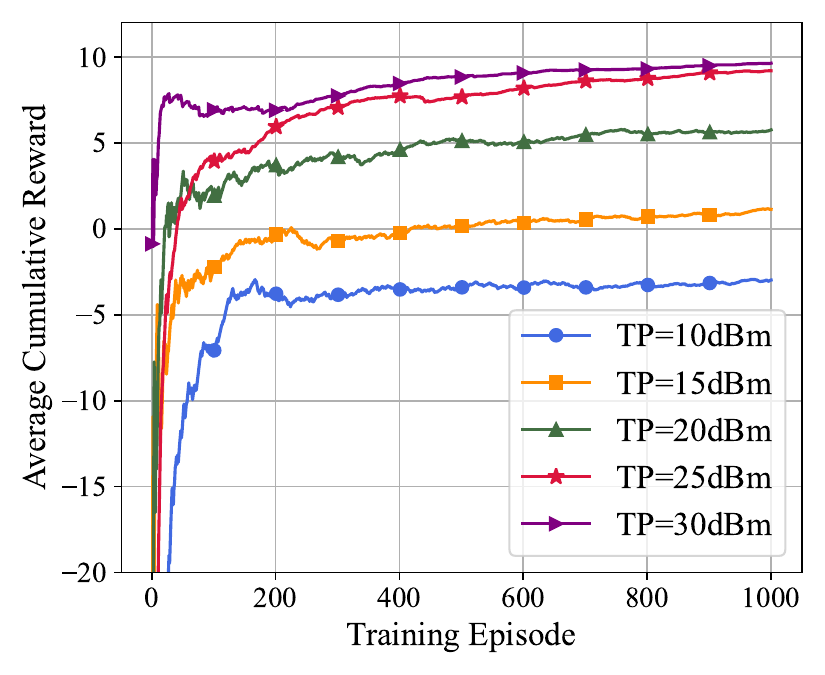}
    \label{vs P_value-episode}
    \hspace{1.2cm}
        \includegraphics[height=2.5in, page=1]{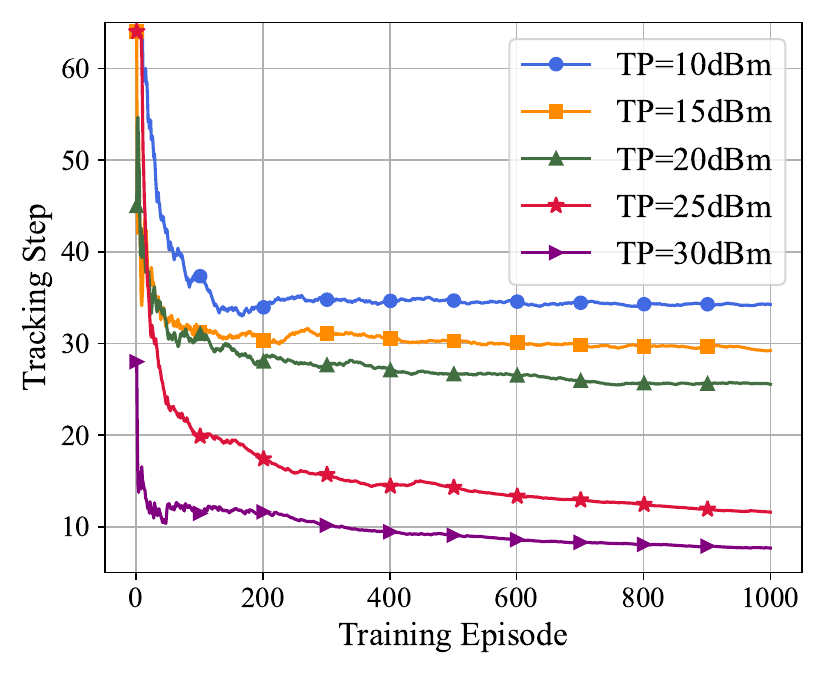}
        \label{vs P_value-step}
    \caption{Convergence behaviors of our DT-assisted DRL under different tranmitting power when $P_\mathrm{FA}=10^{-3}$.}
    \label{vs P_value}
\end{figure*}

Here, we present the convergence behavior of our proposed DT-assisted offline DRL under varying transmitting power levels in Fig. \ref{vs P_value}. The proposed algorithm demonstrates robust stability across transmitting power levels ranging from 10 dBmW to 30 dBmW. Even at low transmitting power, the algorithm requires approximately 30 steps to identify the optimal beam. When the transmitting power increases to 30 dBmW, the algorithm identifies the optimal beam in an average of fewer than 10 steps.

\begin{figure}[t]
	\centering
	\includegraphics[height=2.4in, page=1]{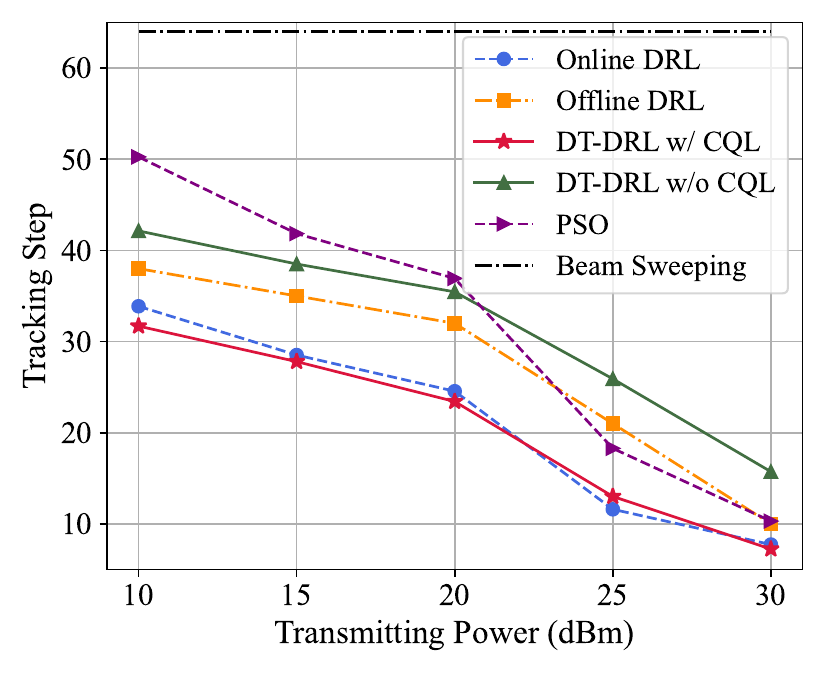}
	\caption{Comparisons of tracking steps between our DT-assisted DRL and other methods under different transmitting power.}
    \label{vs P_value line chart}
\end{figure}

To compare the performance of different methods, we train each DRL algorithm for 1000 episodes in varying transmitting power environments and plot the number of beam tracking steps required to achieve the desired detection performance in Fig. \ref{vs P_value line chart}. Different from Fig. \ref{vs other methods} and Fig. \ref{vs P_value}, the results of offline DRL and DT-assisted DRL are obtained by employing the well-trained model parameters, after complete iterations, into the real environment. It can be observed that all methods (excluding beam sweeping) show a decreasing trend in tracking steps as transmission power increases, indicating improved beam tracking efficiency with higher signal power. Consistent with previous findings, the gap between offline DRL and DT-DRL with CQL highlights that the DT module offers a significant advantage in minimizing tracking steps. Similarly, the designed CQL function proves crucial when comparing DT-DRL with and without CQL. Although the PSO algorithm performs well under high SNR conditions, its performance degrades significantly as SNR decreases. In contrast, our proposed algorithm exhibits strong robustness and consistently achieves remarkable performance, particularly at low SNR levels, demonstrating superior resilience to uncertainties such as environmental noise.

\subsection{Performance under Different CFAR Constraints}
\begin{figure*}[htbp]
	\centering
	\subfloat[$P_\mathrm{FA}=10^{-1}$]{
		\includegraphics[height=1.35in, page=1]{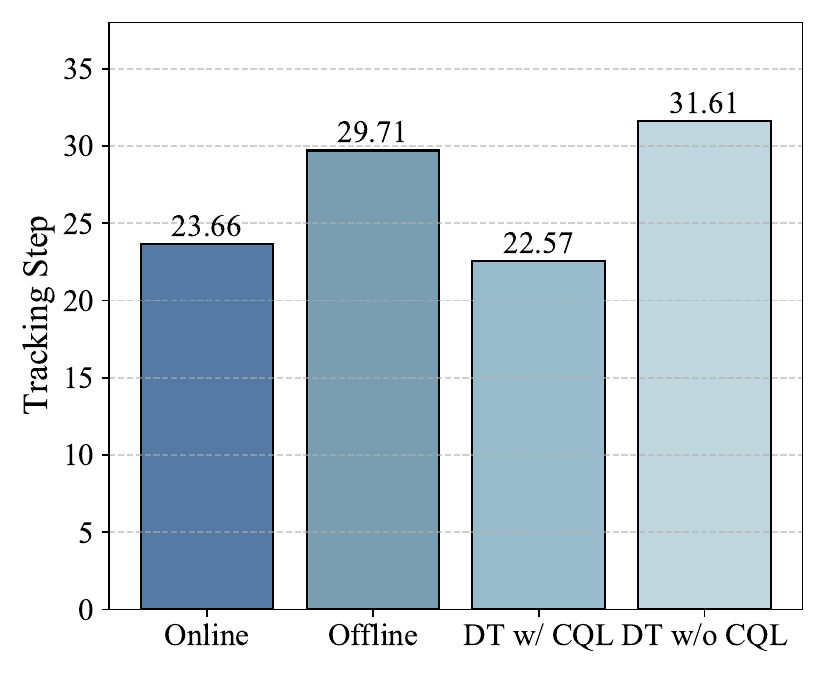}}
	\subfloat[$P_\mathrm{FA}=10^{-2}$]{
		\includegraphics[height=1.35in, page=1]{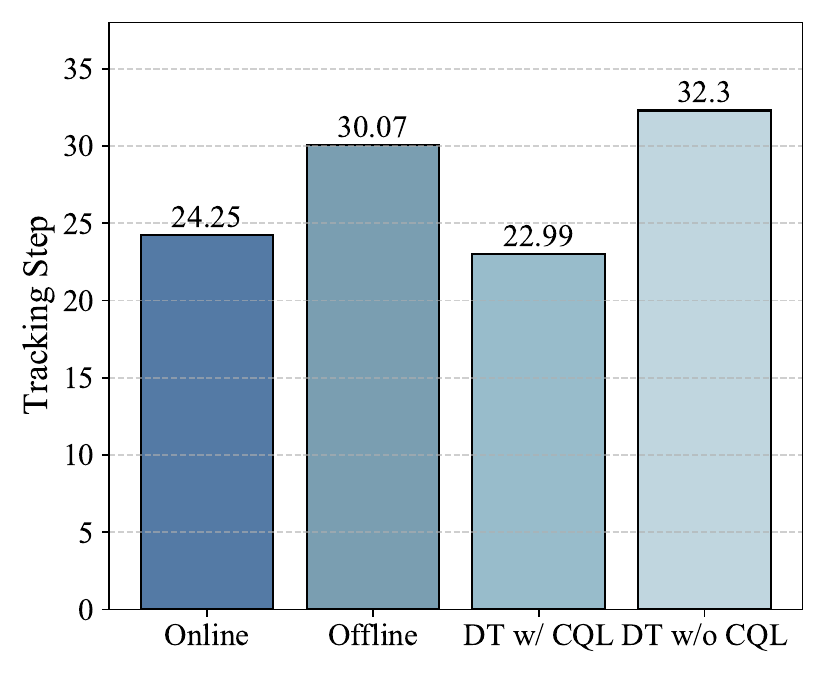}}
	\subfloat[$P_\mathrm{FA}=10^{-3}$]{
		\includegraphics[height=1.35in, page=1]{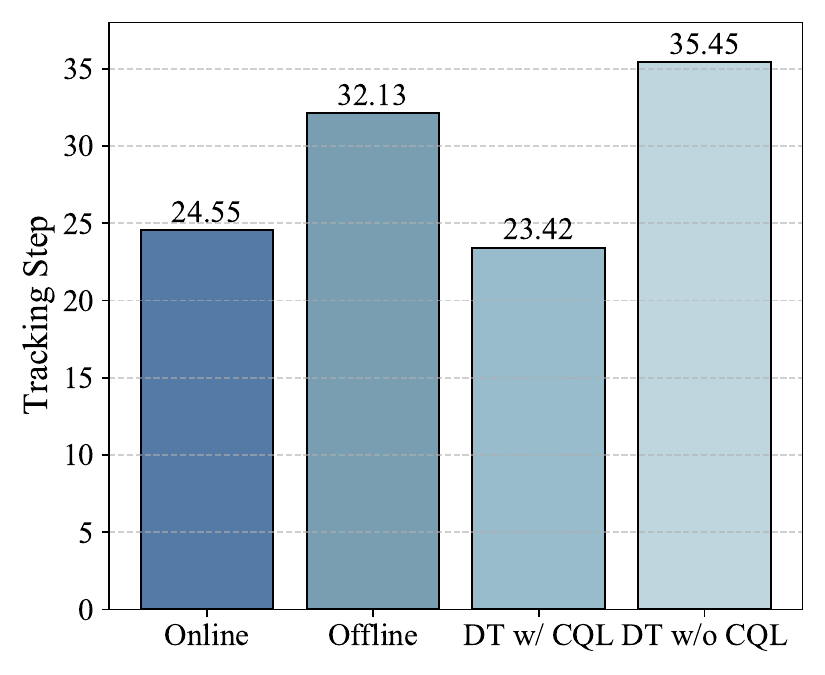}}
        \subfloat[$P_\mathrm{FA}=10^{-4}$]{
		\includegraphics[height=1.35in, page=1]{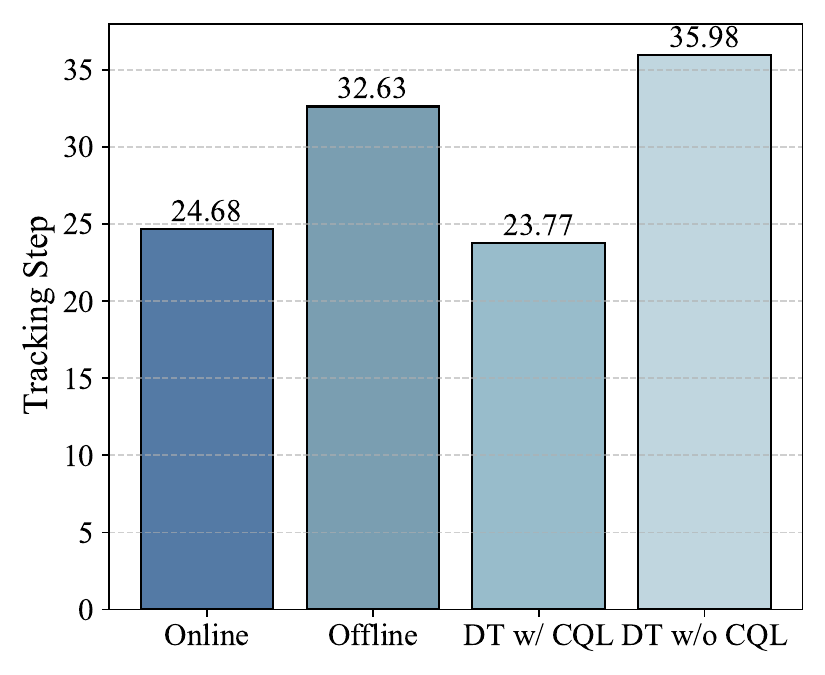}}
	\caption{Comparisons of tracking steps between our DT-assisted DRL and other methods under different CFAR constraints.}
    \label{vs P_FA}
\end{figure*}

Fig. \ref{vs P_FA} presents a comparative analysis of tracking steps for different beamforming strategies under varying CFAR constraints, denoted by values of $P_\mathrm{FA}$. The four subfigures correspond to $P_\mathrm{FA}=10^{-1},10^{-2},10^{-3},$ and $10^{-4}$, illustrating the performance of online DRL, offline DRL, DT-DRL with CQL, and DT-DRL without CQL, respectively. It can be observed that, as $P_\mathrm{FA}$ decreases, tracking steps tend to increase across all methods, indicating a direct impact of CFAR constraints on beamforming performance and a trade-off between detection reliability and tracking efficiency. Stricter false alarm control reduces detection sensitivity, necessitating more steps for beam alignment. Furthermore, DT-DRL with CQL maintains relatively fewer tracking steps compared to DT-DRL without CQL, confirming that incorporating CQL enhances the efficiency of DT-assisted DRL models. Offline DRL requires more tracking steps than DT-DRL with CQL, underscoring the benefits of the DT module. Finally, DT-DRL with CQL proves to be the most effective approach under all CFAR constraints, reinforcing the advantages of our proposed framework for adaptive beamforming.

\subsection{Performance at Different Velocities of The Moving Target}
\begin{figure}[t]
	\centering
	\includegraphics[height=2.0in, page=1]{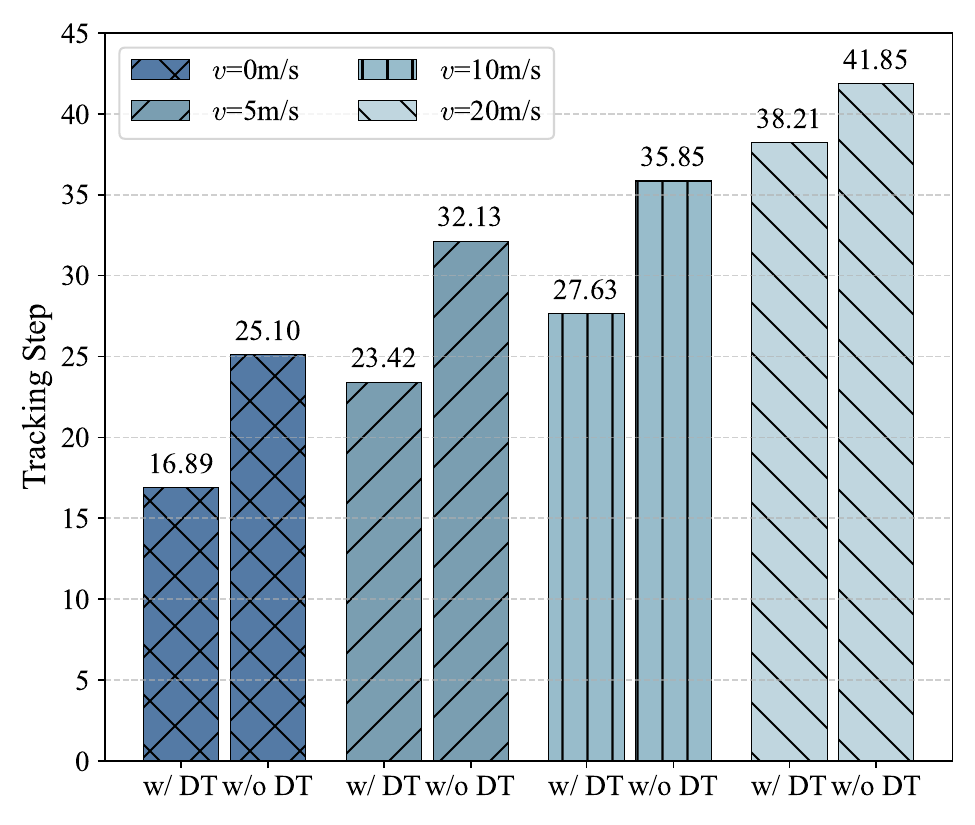}
	\caption{Comparisons of tracking steps of offline DRL with and without DT assistance under different target velocities.}
        \label{vs velocity}
\end{figure}

Taking a step further, Fig. \ref{vs velocity} illustrates the impact of dynamic environmental changes on beamforming strategies, with target velocity changing from $v=0$ to $v=20$m/s. The bars represent the number of tracking steps required for beam alignment. It can be seen that as velocity increases, the number of tracking steps required for beam alignment rises in both DT-assisted and non-DT-assisted cases, reflecting the greater challenge of tracking highly mobile targets. However, offline DRL with DT consistently outperforms its counterpart without DT, requiring fewer tracking steps across all velocity levels. This improvement is attributed to the DT's ability to enhance beam tracking performance through augmented data support. By enriching the diversity of state-action pairs, the DT effectively expands the training domain beyond the original data. Such a capability allows the agent to overcome the limitations of historical data and to generate plausible transition samples corresponding to previously unseen target trajectories or higher target velocities. The performance gap widens as velocity increases, emphasizing the crucial role of DT under dynamic conditions. Without DT, the system struggles to adapt efficiently, leading to a sharper increase in tracking steps. These findings suggest that DT-assisted offline DRL excels in fast-changing environments, where rapid adjustments are essential for maintaining effective beam alignment.

\section{Conclusion}
This paper investigated the problem of joint target detection and proposed a novel beam selection method for cell-free ISAC systems. We employed offline DRL based on a dueling DDQN network and meticulously designed reward shaping function and sinusoidal embedding to enable effective beam selection while ensuring adherence to predetermined false alarm rates and target detection threshold. To improve the learning efficiency of the offline DRL agent, we proposed a DT-assisted framework based on cGAN to enrich the data diversity and incorporated the CQL function to address the overestimation problem. Numerical results demonstrate that our proposed method offers significant advantages in the beam selection task, reducing the number of interactions by 80\% while maintaining performance comparable to online DRL. Furthermore, our proposed method exhibits robustness across various transmitting power levels, false alarm rate probabilities, and target velocities.

In future works, we plan to exploit communication signals to assist target detection and enable joint ISAC optimization. Multi-agent learning frameworks will be investigated to address the exponential growth of the action space as the number of users increases. Moreover, integrating DT with multi-agent DRL represents another promising direction for enhancing system scalability and coordination efficiency.

\ifCLASSOPTIONcaptionsoff
\newpage
\fi

\bibliographystyle{IEEEtran}
\normalem
\bibliography{citations.bib}

\end{document}